\documentclass[a4paper,11pt]{article}
\pdfoutput=1 % if your are submitting a pdflatex (i.e. if you have
\usepackage{jcappub} % for details on the use of the package, please
\usepackage[T1]{fontenc} % if needed

\def\beq{\begin{equation}}
\def\eeq{\end{equation}}
\newcommand{\bea}{\begin{eqnarray}}
\newcommand{\eea}{\end{eqnarray}}

\def\SU{\text{SU}}

\newcommand{\cA}{{\mathcal A}}

\newcommand{\cL}{{\mathcal L}}
\newcommand{\cH}{{\mathcal H}}

\title{\boldmath Non-singular black holes and the Limiting Curvature Mechanism: \\A Hamiltonian perspective}

\author[a]{J. Ben Achour,\note{Corresponding author.}}
\author[b]{F. Lamy,}
\author[b,c,d]{H. Liu}
\author[b,d]{and K. Noui}

\affiliation[a]{Center for Field Theory and Particles Physics, Fudan University\\200433 Shanghai, China}
\affiliation[b]{Laboratoire Astroparticule et Cosmologie, CNRS, Universit\'e Paris Diderot 7\\75013 Paris, France}
\affiliation[c]{Centre de Physique Th\'eorique (UMR CNRS 7332), Universit\'e d'Aix-Marseille et de Toulon\\ 13288 Marseille, France}
\affiliation[d]{Laboratoire de Math\'ematiques et Physique Th\'eorique (UMR CNRS 7350), Universit\'e Francois Rabelais, Parc de Grandmont\\ 37200 Tours, France}

\emailAdd{jibrilbenachour@gmail.com}
\emailAdd{lamyf@apc.in2p3.fr}
\emailAdd{liuh@apc.in2p3.fr}
\emailAdd{karim.noui@lmpt.univ-tours.fr}

\abstract{We revisit the non-singular black hole solution in (extended) mimetic gravity 
with a limiting curvature from a Hamiltonian point of view. We introduce a parameterization of the phase space which allows 
us to describe fully the Hamiltonian structure of the theory. We write down the equations of motion that we solve in the regime deep 
inside the black hole, and we recover that the black hole has no singularity, due to the limiting curvature mechanism. 
Then, we study the relation between such black holes and effective polymer black holes which have been introduced in the context of loop quantum gravity. 
As expected, contrary to what happens in the cosmological sector, mimetic gravity with a limiting curvature fails to reproduce 
the usual effective dynamics of  spherically symmetric loop quantum gravity which are generically not covariant. Nonetheless, we exhibit a theory in the class of extended mimetic gravity whose dynamics reproduces the general shape of the effective corrections of spherically symmetric polymer models, but in an undeformed covariant manner. These covariant effective corrections are found to be always metric dependent, i.e. within the $\bar{\mu}$-scheme, underlying the importance of this ingredient for inhomogeneous polymer models. In that respect, extended mimetic gravity can be viewed as an effective covariant theory which naturally implements a covariant notion of point wise holonomy-like corrections. The difference between the mimetic and polymer Hamiltonian formulations provides us with a guide to understand the deformation of covariance in inhomogeneous polymer models. }

\begin{document}
\maketitle
\flushbottom

\newpage

\section{Introduction}

\subsection{Motivations: mimetic gravity as a guide to derive a covariant effective loop regularization}

Effective symmetry reduced polymer models of quantum geometry provide us with an interesting framework to investigate the deviations from General Relativity induced by the quantum nature of the geometry. These models are inspired by Loop Quantum Gravity, from which the regularization and quantization technics are borrowed. In its canonical form, the regularization step is performed at the level of the phase space, prior to quantization, such that the canonical variables of GR as well as the first class constraints encoding the gauge symmetry are modified.  
From this perspective, two natural questions appear regarding the consistency of the approach:
\begin{itemize}
\item i)  First, does the classical regularization generate anomalies or deformations in the algebra of first class constraints, i.e. in the Dirac's hypersurface deformation algebra (DHDA) ? 
\item ii) Second, can one obtain an effective covariant action which reproduces, in the symmetry reduced sector under consideration, the effective quantum corrections implemented at the hamiltonian level ?
\end{itemize}
Naturally, these two questions are intimately related since no covariant action could reproduce effective hamiltonians corrections if they generate deformation of the Dirac's algebra, as it is the case in most inhomogeneous polymer models \cite{Bojowald:2015zha, Bojowald:2015sta, Bojowald:2016hgh, Bojowald:2016vlj, Bojowald:2016itl, Tibrewala:2013kba, Tibrewala:2012xb}.

In the homogeneous cosmological sector, where there is no spatial diffeomorphism, the question i) of anomaly freedom naturally disappears and the DHDA is trivial. Therefore the regularization is not constrained by any anomaly free conditions. One can thus derive en effective cosmological phase space by introducing the effective corrections inspired by LQG, and study the modified system. The resulting effective quantum cosmology, i.e. Loop Quantum Cosmology, describes a bouncing universe with a consistent semi-classical limit. One can quantize this system using again technics from LQG and show that for sufficiently semi-classical states, the quantum dynamics coincides with the effective classical dynamics, reinforcing the consistency of the approach at least in this cosmological sector. Moreover, the effective dynamics of LQC can be derived from a covariant scalar tensor lagrangian belonging to the extended mimetic gravity family \cite{Liu:2017puc,Bodendorfer:2017bjt}. Hence, in this cosmological case, question ii) can be answered  in the affirmative.

A natural question is whether the successful construction of LQC, implementing the (partial) loop regularization with its associated effective (mimetic) action, can be generalized to inhomogeneous gravitational systems. For example, does it extend to the inhomogeneous spherically symmetric case ? Does the solution of a regular black hole interior found in \cite{Chamseddine:2016ktu} (from the Chamesddine-Mukhanov's lagrangian which reproduces the effective dynamics of LQC in the cosmological sector) is related to the effective polymer black hole models \cite{Gambini:2013ooa, Bojowald:2015zha, BenAchour:2018khr, Bojowald:2018xxu} studied so far ?

As explained in more details in Sec.I.D, a first obstacle in generalizing the construction of LQC and its relation the extended mimetic theory found in \cite{Liu:2017puc,Bodendorfer:2017bjt}, is that when going to inhomogeneous situations, such as inhomogeneous cosmology, spherical symmetry as well as Gowdy system, point i) is no more satisfied. When no local degrees of freedom are involved, the Dirac's algebra has been shown to be deformed by holonomy corrections, while in presence of local degrees of freedom, the DHDA is anomalous. These results call therefore for a generalization of the current loop regularization used in inhomogeneous polymer models. Yet, as it stands, the current situation prevents from the start to derive an covariant effective action for these inhomogeneous models as done for homogeneous LQC in \cite{Liu:2017puc,Bodendorfer:2017bjt}. 

In this work, we adopt to following strategy: we assume that extended mimetic gravity can indeed provide a consistent effective theory of polymer quantum gravity and study its symmetry reduction to spherical geometry. We propose a general procedure which, given some holonomy corrections, allow to derive the associated effective action belonging to the extended mimetic gravity family. This will provide a guide in investigating the issue of covariance in inhomogeneous polymer black hole models.

Having presented our motivations, let us now review briefly the extended mimetic gravity framework and the limiting curvature mechanism which will be a corner stone of our work. In order to be self complete, we also include a brief review of the recent results concerning the above mentioned issue of covariance in inhomogeneous effective polymer models. In the following, we shall use Planck units, i.e $G=c= 1$.

\subsection{Equivalent formulations and relation to DHOST theories}

Mimetic gravity \cite{Chamseddine:2013kea}  is a theory of modified gravity which belongs to the family of
scalar-tensor theories.   There have been numerous generalizations of mimetic gravity, but the simplest (and original) theory is 
simply defined by the Einstein-Hilbert action for a metric $\tilde{g}_{\mu\nu}$
\bea
\label{mimintro}
\tilde S[\tilde g_{\mu\nu},\phi] \; = \; \int d^4x \sqrt{-g} \, {\cal R} \quad
\text{where} \quad g_{\mu\nu} \equiv -\tilde X \tilde g_{\mu\nu} \quad \text{and} \quad \tilde X\equiv \tilde g^{\mu\nu}\phi_\mu \phi_\nu \, .
\eea
where we use the standard notation $\phi_{\mu} = \partial_{\mu} \phi$ and $X= \phi^{\mu} \phi_{\mu}$. The metrics $g_{\mu\nu}$ and $\tilde{g}_{\mu\nu}$ are related by a non-invertible
conformal transformation: if one can uniquely define $g_{\mu\nu}$ from ${\tilde{g}}_{\mu\nu}$, the reverse is not true.
Indeed, the metric $\tilde{g}_{\mu\nu}$ is obviously defined up to a conformal transformation only.

The Euler-Lagrange equations derived from this action \eqref{mimintro} are rather different from the 
Einstein equations even if the action reduces to  the Einstein-Hilbert action for $\tilde g_{\mu\nu}$. To make this distinction clear, let us remark that 
the metric $g_{\mu\nu}$ satisfies the constraint $X\equiv g^{\mu\nu}\phi_\mu\phi_\nu = -1$, which allows us to reformulate the action \eqref{mimintro} equivalently as
\bea
\label{secondform}
S[g_{\mu\nu},\phi,\lambda] \; = \; \int d^4x \sqrt{- g} \, \left[ {\cal R} + \lambda (X+1)\right] \, ,
\eea
where $\lambda$ is a new dynamical variable. It is easy to show that these two
actions \eqref{mimintro} and \eqref{secondform} 
produce two equivalent sets of equations of motion. Thus, mimetic gravity appears to be equivalent to gravity coupled
to the scalar fields $\phi$ and $\lambda$. A careful analysis shows that the theory propagates only one scalar degree of freedom in 
addition to the usual two tensorial modes (see \cite{Chaichian:2014qba} but also \cite{Langlois:2015skt} for a Hamiltonian analysis of 
general higher-order scalar-tensor theories). 
Notice that, in its first formulation \eqref{mimintro}, the theory is clearly conformally invariant: the metric $\tilde{g}_{\mu\nu}$ and thus the action itself is invariant under the transformation $g_{\mu\nu}\mapsto \Omega(x) g_{\mu\nu}$ where $\Omega$ is an arbitrary function
on the space-time. Hence, the conformal mode of the metric becomes dynamical, and this is the deep reason why the theory 
propagates one more degree
of freedom than general relativity. Since mimetic gravity has been introduced, it has been studied extensively 
mainly in the context  of cosmology (see \cite{Sebastiani:2016ras} and references therein).

\medskip

Starting from the first formulation \eqref{mimintro}, one can easily expand the Ricci scalar ${{\cal R}}$ and the measure $\sqrt{-g}$ in terms of $\tilde g_{\mu\nu}$  and $\phi$, and  show that the action contains higher derivatives of the scalar field according to
\bea
\label{DHOSTmim}
\tilde S[\tilde g_{\mu\nu},\phi] \; = \; - \int d^4x \sqrt{-\tilde g} \left[ \tilde{X} \tilde{{\cal R}} + \frac{6}{\tilde X} \tilde{g}^{\mu\rho} \tilde g^{\nu\alpha}\tilde g^{\sigma\beta}   \, \phi_{\mu\nu} \phi_{\rho\sigma} 
\phi_\alpha \phi_\beta \right] \, ,
\eea
where $\tilde {\cal R}$ is the Ricci scalar associated to the metric $\tilde g_{\mu\nu}$ and $\phi_{\mu\nu} = \nabla_{\mu} \partial_{\nu} \phi$.
In that way, mimetic gravity
can be viewed explicitly as a higher-order scalar-tensor theory 
which is moreover degenerate in the sense of \cite{Langlois:2015cwa}. Thus, it belongs to the
family of Degenerate Higher-Order Scalar-Tensor (DHOST) theories \cite{Langlois:2015skt,Crisostomi:2016tcp,Crisostomi:2016czh,Achour:2016rkg,BenAchour:2016fzp}, which confirms that it propagates (up to) only three degrees of freedom: one scalar and two tensorial modes. The fact that \eqref{DHOSTmim} is degenerate is closely related to the conformal invariance of the theory. This extra symmetry insures the presence of an extra (first class) constraint in the theory (in addition to the usual Hamiltonian and vectorial constraints) which makes the theory degenerate.  Notice that the conformal invariance has been somehow broken in the second formulation of the theory
\eqref{secondform} at the cost of bringing in the scenario an extra scalar degree of freedom $\lambda$. 

\subsection{Limiting curvature hypothesis}
The initial physical motivation for considering the theory \eqref{mimintro} has been to propose an alternative to
  cold dark matter in the universe.
Actually, it reproduces exactly the results of a model introduced earlier by Mukohyama in \cite{Mukohyama:2009mz}. 
Later on, the original proposal of  \cite{Chamseddine:2013kea}  has been extended, in adding a potential $V(\phi)$ for the scalar field
in \eqref{mimintro} and has been shown to provide potentially interesting models 
for both the early universe and the late time cosmology \cite{Chamseddine:2014vna}. Finally, more recently, mimetic gravity has been
applied to construct non-singular cosmologies and non-singular black holes in  \cite{Chamseddine:2016uef,Chamseddine:2016ktu}. 
Physically, the idea is very simple
and consists in finding higher-order scalar-tensor Lagrangians (in the family of generalized mimetic gravity) 
in such a way that their Euler-Lagrange equations impose an upper limit on the Ricci scalar ${\cal R}$:
this is called the limiting curvature hypothesis. If this is the case,  ${\cal R}$ never diverges
and one could expect to resolve in that way some divergences which appear in classical  general relativity.

Such a hypothesis can be implemented concretely and, in some situations, it is
 sufficient to transform the original cosmological singularity into a non-singular bounce in
the context of homogeneous and isotropic space-times \cite{Chamseddine:2016uef}, and also to remove the black-hole singularity
in the context of spherically symmetric space-times.  The limiting curvature hypothesis  was extended very recently in \cite{Yoshida:2017swb} 
to construct theories whose equations of motion
impose upper limits not  only on the Ricci scalar but also on any invariant constructed from the Riemann tensor 
${\cal R}_{\mu\nu\rho\sigma}$ such as the Ricci tensor squared ${\cal R}_{\mu\nu}{\cal R}^{\mu\nu}$, the Weyl tensor squared ${C^2}=C_{\mu\nu\rho\sigma}C^{\mu\nu\rho\sigma}$, or invariants involving derivatives of the Riemann.
However, as the authors pointed out in \cite{Yoshida:2017swb}, these extended Lagrangians contain higher-order derivatives of the metric 
(which is not the case for the original mimetic gravity)  and presumably suffer from Ostrogradski instabilities  generically \cite{Crisostomi:2017aim}. 
Moreover, it should be stress that up to now, no investigation on the stability of such models in the spherically symmetric framework has been performed. As a result, one should consider these theories of gravity with a limiting curvature as effective descriptions which are physically valuable up to some energy scale only.

\subsection{Towards an effective description of polymer black hole in Loop Quantum Gravity?}
Interestingly, as mentioned earlier, it was realized in \cite{Liu:2017puc,Bodendorfer:2017bjt} that the original theory of mimetic gravity with a limiting curvature introduced in \cite{Chamseddine:2016uef} reproduces exactly 
the effective Loop Quantum Cosmology bounce (LQC) \cite{Ashtekar:2011ni} 
as a homogeneous and isotropic solution\footnote{Notice
that a very similar construction has been considered much earlier in \cite{Helling:2009ia} in an attempt to reproduce the effective dynamics
of LQC from an $f(R)$ theory.}. 
Naturally, such an observation raises the question of  whether one could consider mimetic gravity with a limiting curvature
as an effective description (up to some energy scale) of the full theory of Loop Quantum Gravity (LQG), even  out of the cosmological sector.  

Nonetheless, there are at least two obstacles for such a scenario to be possible. On the one hand, LQC is not a subsector of the full theory of LQG but rather a homogenous symmetry reduced model which is quantized importing the technics of the full LQG theory. Despite some preliminary results concerning the embedding of LQC into LQG \cite{Alesci:2016gub, Alesci:2015nja, Bodendorfer:2015qie, Livine:2011up,  BenAchour:2016ajk}, this aspect remains 
up to now poorly understood. Therefore the relation established between LQC and mimetic gravity does not say much  a priori 
about an eventual link between LQG and mimetic gravity. 
On the other hand, even without considering the embedding of LQC into LQG, one could wonder whether the result obtained in \cite{Liu:2017puc,Bodendorfer:2017bjt} 
can be generalized to less symmetric situations, such as spherically symmetric \cite{Gambini:2013ooa} or Gowdy loop models \cite{deBlas:2017goa}. Such symmetry reduced models exhibit non perturbative inhomogeneities, and as such, are highly non trivial to quantize using the LQG technics. The reason is that, contrary to the homogenous cosmological sector, the invariance under spatial diffeomorphisms survives the symmetry reduction. Therefore, the loop quantization of such inhomogenous models faces the additional difficulty of keeping the spatial covariance (in the sense of the symmetry reduced model)
in the quantization. More precisely, the issue of the covariance takes the form of requiring (at the quantum level) an anomaly free  Dirac's hypersurface deformation algebra (DHDA), which is the algebra of the first class constraints generating the invariance under diffeomorphisms. 
Classically, this algebra is generated by the Hamiltonian constraint $\cH$, the vectorial constraints $D_a$ satisfying 
\bea
&& \{ D[U^a], D[V^b]\} = D[\cL_{{U}} V^a] \, ,\quad \{ D[U^a], \cH[N]\} =  - \cH[\cL_{{U}} N]  \, , \label{diffintro}\\
&& \{ \cH[N], \cH[M]\} =  D[ \gamma^{ab} \big{(} N \partial_b M - M \partial_b N\big{)}]  \, , \label{Ham}
\eea
where we have used the following standard notations: $U=U^a \partial_a$ and $V=V^a\partial_a$ are spatial vectors, $N$ and $M$ are scalars,  
$D[U^a]$ and $\cH[N]$ are the smeared vectorial and Hamiltonian constraints respectively, $\gamma^{ab}$ is the inverse spatial metric and ${\cL}_{{U}}$
denotes the usual Lie derivative along the vector $U$. In the quantum theory, this algebra is modified. Indeed, introducing point-wise holonomy corrections to regularize the first class constraints, as required by the loop quantization procedure, has very important consequences. The fate of the DHDA under the introduction of the holonomy corrections has been studied recently in a series of papers \cite{Bojowald:2015zha, Bojowald:2015sta, Bojowald:2016hgh, Bojowald:2016vlj, Bojowald:2016itl}. A generic result is that for system having no local degrees of freedom, such as vacuum spherically symmetric gravity \cite{Bojowald:2015zha}, or Gowdy model with local rotational symmetry \cite{Bojowald:2015sta}, the DHDA exhibits a deformed notion of covariance 
in the sense that  \eqref{Ham} is replaced by
\bea
\label{def}
 \{ \cH[N], \cH[M]\} =  \beta{(\tilde{K})}D[ \gamma^{ab} \big{(} N \partial_b M - M \partial_b N\big{)}]   \, ,
\eea
where $\beta(\tilde{K})$ is a ``deformation", called sometimes the ``sign change deformation",  which depends generically on the homogeneous component of the extrinsic curvature $\tilde{K}$.
For systems having local degrees of freedom, the situation is even worse since the DHDA has been shown not to be a closed algebra anymore. 

All this seems to lead us to the conclusion that it is not possible to provide an effective description of such loop quantized phase space using a covariant scalar tensor theory such as \cite{Chamseddine:2016uef, Yoshida:2017swb}, since these theories share the same (undeformed) DHDA as classical General Relativity (augmented with a scalar degree of freedom). A first notorious complication in such model is that the quantum corrections are implemented at the Hamiltonian level and going back to a Lagrangian formulation is often very complicated (see \cite{Bojowald:2011aa} for earlier investigations on this aspect). Nonetheless, the situation is not totally hopeless. There are at least two reasons to think that one can circumvent these apparent obstacles we have discussed above. The first one relies on the observation that the difficulties related to the deformation of the DHDA seems to be inherent to the fact of working with the real Ashtekar-Barbero variables. Recent results have shown that such difficulties disappear when working with the initial self-dual variables, at least in the case of spherical symmetry gravity coupled to matter and unpolarized Gowdy models \cite{BenAchour:2016brs, BenAchour:2017jof}. In this situation, the DHDA keeps indeed its classical form without any deformation even if some holonomy corrections, including a $\bar{\mu}$-scheme, have been taken into account. Therefore, this opens the possibility of describing these self-dual loop symmetry reduced models using a covariant scalar-tensor theory. 
The second one relies on the fact that up to now, the polymer construction of black hole models (as well as inhomogeneous cosmology and Gowdy system) has focused only on the minimal loop regularization, namely the point-wise holonomy corrections. Yet, only few investigations have considered together the additional corrections inherent to LQG: the triad corrections related to the inverse volume regularization, as well as the holonomy of the inhomogeneous component of the connection. Therefore, it is not clear if the deformation (\ref{def}) will not be removed when one takes into account  these additional corrections.

Facing the problem of the deformed notion of covariance in polymer models inspired from LQG, we adopt the following strategy. As shown in \cite{Liu:2017puc,Bodendorfer:2017bjt}, mimetic gravity with a limiting curvature is a covariant scalar-tensor theory which reproduces in the cosmolgical sector the effective dynamics of LQC. Thus, we consider this theory as a potential guide to obtain an effective Lagrangian description of point wise holonomy-like corrections which maintain the covariance beyond the cosmological sector. As a first study, we focus our attention on the spherically symmetric sector and derive its Hamiltonian formulation. As expected, it leads to effective quantum corrections similar to the ones introduced in the polymer framework, but in a covariant manner. The difference between the mimetic effective corrections and the polymer ones provides an interesting guide to understand the lack of covariance of polymer black hole models.
  
\subsection{On the effective description of quantum black holes in LQG}
Before closing this introduction, let us explain more precisely how quantum black holes have been treated in the framework of LQG.
The first descriptions of quantum black holes in LQG were introduced in \cite{Rovelli:1996dv,Ashtekar:1997yu}.  
They allowed to obtain  a complete description of the black hole microstates whose 
 counting is very well-known to reproduce the Bekenstein-Hawking formula at the semi-classical limit. Even though  important issues concerning the role of the Barbero-Immirzi parameter in the counting procedure remain unsolved, (see \cite{Frodden:2012dq, Achour:2014eqa, BenAchour:2016mnn, BenAchour:2014zoub, Geiller:2014eza, Geiller:2012dd, Achour:2013gga, Achour:2013hga} for some proposals to overcome them), this result has been an important success for LQG. Unfortunately, these studies do not say much about the dynamical ``geometry" of the quantum
 black holes in the context of LQG: microstates are described in terms of algebraic structures (intertwiners between representations) and recovering the geometrical content
 of the quantum black holes at the semi-classical limit is a very complicated task which required coarse-graining schemes still under development. One way to attack the problem would be
 to find, exactly as it was done in LQC, an effective description of spherically symmetric solutions and see how quantum gravity effects modifies Einstein equations in this sector. 
 
Such a program, which relies on a polymer quantization of spherically symmetric geometries, was initiated in \cite{Ashtekar:2005qt} and further developed in \cite{Bohmer:2007wi, Modesto:2004wm, Modesto:2006mx, Campiglia:2007pb}. The effective dynamics of the interior black hole was explored in \cite{Corichi:2015xia} and its construction follows very closely what is done in LQC, the black hole interior geometry being a homogenous anisotropic cosmology. However, the treatment of the whole space-time and its inhomogenous exterior geometry remains more challenging. While the loop quantization of the inhomogeneous vacuum exterior geometry was worked out in details in \cite{Gambini:2013ooa, Gambini:2008dy}, an effective theory of this quantum geometry is not yet available (see \cite{Gambini:2008dy} for a review on the quantization of the full spherically symmetric geometry). Moreover, going beyond the vacuum case and including matter has also proved to be very challenging, mainly because the standard holonomy corrections spoil generally the covariance of such models \cite{Bojowald:2015zha, Bojowald:2016vlj}. For this reason, studying an effective inhomogenous gravitational collapse including the quantum corrections from loop quantum gravity is up to now out of reach with the standard techniques. Another way to understand the geometry of quantum 
black holes (at the semi-classical limit) would be to ``guess" the modifications induced by quantum gravity effects in mimicking LQC as it was done in \cite{Rovelli:2014cta} where 
the notion of ``Planck stars" has been introduced. Some phenomenological aspects of these potentially new astrophysical objects have been studied in \cite{Rovelli:2014cta,Barrau:2014hda, Christodoulou:2016vny,DeLorenzo:2014pta}. While the heuristic idea of the Planck stars is fascinating, a quantum theory of spherically symmetric geometry coupled to matter from which such effective description emerged still has to be built.

Here, we investigate an alternative way to study the issue of covariance in polymer models inspired by LQG using mimetic gravity.
We start with  a Hamiltonian analysis of the theory \cite{Chamseddine:2016uef} and we revisit the non-singular black hole solution from a Hamiltonian point of view. 
In particular, we find  the expression of the Hamiltonian constraint which could be interpreted as an effective Hamiltonian constraint for spherically symmetric geometries 
in quantum gravity. However, this expression differs from the ones that have been considered so far in the literature (even though their classical limits obviously agree). Nonetheless, this offers a new regularization scheme to be compared with the one obtained in standard  spherically symmetric loop models \cite{Gambini:2008dy}. One outcome of our approach is that the effective holonomy-like corrections have a natural inbuilt $\bar{\mu}$-scheme, contrary to the holonomy corrections usually introduced in the standard loop treatment \cite{Gambini:2008dy} (see \cite{Tibrewala:2012xb} for an attempt to include phase space dependent holonomy corrections in spherically symmetric polymer models).
Moreover, we show that one can easily modify the initial proposal of  \cite{Chamseddine:2016uef} in order to obtain a theory of mimetic gravity which reproduces
the usual effective Hamiltonian constraint of LQG for spherically symmetric geometries, including again an additional $\bar{\mu}$-scheme. 

\subsection{Layout of the paper}
The paper is organized as follows. In section II, we recall some important aspects of the extended theories of mimetic gravity and we illustrate the limiting curvature mechanism in the context of black holes. In section III, we revisit the non-singular black hole solution described in  \cite{Chamseddine:2016uef}
from a Hamiltonian point of view. In particular, we find a parametrization of the phase space which makes the resolution of the equations of motion
simpler than in the Lagrangian formulation. Furthermore, we exhibit the explicit form of the Hamiltonian constraint for mimetic gravity with a limiting curvature
in the case of a spherically symmetric space-time. In section IV, we compare the Hamiltonian formulation  of this theory of mimetic gravity with those obtained in LQG from a ``loop" regularization of the usual Hamiltonian constraint of gravity. Contrary to what happens in the cosmological sector, the theory of mimetic gravity proposed in  \cite{Chamseddine:2016uef} does not reproduce the effective dynamics of spherically symmetric LQG. However,  we exhibit a theory in the class of extended mimetic gravity whose dynamics reproduces the general shape of the effective corrections  of spherically symmetric polymer models, but in an undeformed covariant manner. In that respect, extended mimetic gravity can be viewed as an effective covariant theory which naturally implements a covariant notion of point wise holonomy-like corrections similar in spirit to the ones used in polymer models. The difference between the mimetic and polymer Hamiltonian formulations provides us with a guide to understand the lack of covariance in inhomogeneous polymer models.

\section{Mimetic gravity and the limiting curvature hypothesis}
In this section, we review some aspects of  extended theories of mimetic gravity \cite{Takahashi:2017pje, Langlois:2017mxy}. Extended mimetic gravity generalizes the original proposal of \cite{Chamseddine:2013kea}: 
it is still  conformally invariant and propagates generically only one scalar in addition to the usual two tensorial modes. Then, we recall how to implement  the limiting curvature hypothesis in this context and  how the proposal of \cite{Chamseddine:2016uef} leads indeed to a non-singular black hole solution.

\subsection{Extended mimetic gravity}
Let us start by discussing the original theory of mimetic gravity defined by the action \eqref{secondform} 
\bea\label{mimetic action}
S[g_{\mu\nu},\phi,\lambda] \; \equiv \; \int d^4x \, \sqrt{-g} \, \left[ \frac{1}{2} {\cal R} + \lambda( g^{\mu\nu} \phi_\mu \phi_\nu \pm 1)\right]\, ,
\eea
where the sign $(\pm 1)$ is for the moment arbitrary and will be fixed hereafter: the $+$ (resp. $-$) sign implies that $\phi^\mu$ are the 
components of a timelike (resp. spacelike) vector.
The dynamical variables are the metric $g_{\mu\nu}$ with signature $(-1,+1,+1,+1)$, 
the scalar field $\phi$ and the extra variable $\lambda$. 

\subsubsection{Equations of motion}
The equations of motion are easily computed, and the Euler-Lagrange equations for
$\lambda$, $\phi$ and $g_{\mu\nu}$ are respectively given by
\bea\label{eom mimetic}
X  \pm 1 \; = \; 0 \, , \qquad
\nabla_\mu(\lambda \phi^\mu) \; = \; 0 \, , \qquad G_{\mu\nu} \; = \; T_{\mu\nu} \, ,
\eea
where the stress-energy tensor for the ``mimetic" matter field is
\bea
T_{\mu\nu} \; = \; -2 \lambda \phi_\mu \phi_\nu +  \lambda(X \pm 1) g_{\mu\nu} \, .
\eea
Let us recall that we used the same notations as in the introduction for the gradient $\phi_\mu \equiv \nabla_\mu \phi$ and the kinetic energy 
$X\equiv \phi^\mu \phi_\mu$. The trace 
of the last equation in \eqref{eom mimetic} allows us to express the variable $\lambda$ in terms of the Ricci scalar
according to
\bea
\lambda \; = \; \pm \frac{1}{2} {\cal R} \, .
\eea
Substituting this expression for $\lambda$ into the first two equations in \eqref{eom mimetic} leads to the following 
system of equations for the scalar field and the metric
\bea
\nabla_\mu({\cal R} \phi^\mu) \; = \; 0 \, , \qquad 
G_{\mu\nu} \pm {\cal R} \phi_\mu \phi_\nu \, = \; 0 \,  .
\eea
Notice that the first equation (for the scalar field) is not independent from Einstein equations. Indeed, the conservation
of the Einstein tensor $\nabla^\mu G_{\mu\nu}$ together with the mimetic condition necessarily imply that
\bea
0 \; = \;  \nabla^\mu G_{\mu\nu}  =   \nabla^\mu({\cal R} \phi_\mu \phi_\nu) =  \nabla^\mu({\cal R} \phi_\mu) \phi_\nu +
 {\cal R} \phi^\mu \phi_{\mu\nu}  =  \nabla^\mu({\cal R} \phi_\mu) \phi_\nu \Longrightarrow 
 \nabla_\mu({\cal R} \phi^\mu)  =  0  \, .
\eea
Hence, the equations of motion \eqref{eom mimetic} are equivalent to 
\bea
G_{\mu\nu} \pm {\cal R} \phi_\mu \phi_\nu \, = \; 0 \, , \quad
X \pm 1 \, = \, 0 \, , \quad
\lambda \; = \; \pm \frac{1}{2} {\cal R} \,.
\eea
These equations have been solved for cosmological space-times in \cite{Chamseddine:2013kea}.

\subsubsection{Generalization: extended mimetic gravity}
The mimetic action \eqref{mimetic action} can be generalized to the form \cite{miminprep}
\bea\label{general mimetic action}
S[g_{\mu\nu},\phi,\lambda] \; \equiv \; \int d^4x \, \sqrt{-g} \, \left[ \frac{f(\phi)}{2} \, {\cal R} + 
\, L_\phi(\phi,\chi_1,\cdots,\chi_p) \, + \, \lambda( X \pm 1)\right]\, ,
\eea
where $f$ is an arbitrary function of $\phi$, $L_\phi$ depends on $\phi$ and $\chi_n$ which are variables 
constructed with  second derivatives of the scalar field according to
\bea
\label{chin}
\chi_n \; \equiv \; \text{Tr}([\phi]^n) \; \equiv \; \sum_{\mu_1,\cdots,\mu_n} \phi_{\mu_1}^{\mu_2} \, \phi_{\mu_2}^{\mu_3} \cdots \phi_{\mu_{n-1}}^{\mu_n} \, \phi_{\mu_n}^{\mu_1} \, .
\eea
Here, we have used the notation $[\phi]$ for the matrix whose coefficients are $[\phi]_{\mu\nu}\equiv \phi_{\mu\nu}$. Indices are lowered and raised 
by the metric and its inverse.
One can show that this extended action defines  the most general mimetic gravity like theory \cite{Takahashi:2017pje, Langlois:2017mxy} which propagates at most three degrees of freedom (one scalar in addition to the usual two tensorial modes). Notice that mimetic gravity has recently been generalized in \cite{Gorji:2017cai} to actions with higher derivatives of the metric (which propagate more than three degrees of freedom out of the unitary gauge).
Due to the mimetic condition $X \pm1 = 0$, any $X$ dependency in 
$f$ or $L_\phi$ can be removed. More precisely, as it was shown in \cite{Langlois:2017mxy},  
if one starts with an action  \eqref{general mimetic action} where $f$ and $L_\phi$
depend also on $X$ (and eventually its derivatives $\partial_\mu X$), the associated equations of motion are equivalent to the equations of motion obtained
from the same action where $f$ and $L_\phi$ are evaluated to $X=\mp 1$ (and eventually $\partial_\mu X=0$). 

The Euler-Lagrange equations for \eqref{general mimetic action} can be easily obtained in full generality. But, for simplicity, we assume that
$f$ is a constant (and thus independent of $\phi$) which  can be fixed to $f=1$. Deriving the action with respect to $\phi$ and $g_{\mu\nu}$
respectively leads to the equations
\bea\label{general mimetic eom}
\frac{\partial L_\phi}{\partial \phi} - 2 \nabla^\mu(\lambda \phi_\mu)+  \sum_{n=1}^p {n} \,\nabla^{\mu \nu}  \left( [\phi]^{n-1}_{\mu\nu} \; 
\frac{\partial L_\phi}{\partial \chi_n}  \right) \; = \; 0 \;  \quad \text{and} \quad
G_{\mu\nu} \; = \; T_{\mu\nu} \, ,
\eea
where $[\phi]^n$ is the power $n$ of the matrix $[\phi]$ with the convention $[\phi]^0_{\mu\nu}\equiv g_{\mu\nu}$, and now the stress-energy tensor reads 
\bea
T_{\mu\nu} &= & -2 \lambda \phi_\mu \phi_\nu +  \lambda(X \pm 1) g_{\mu\nu} \, + \, T^{(\phi)}_{\mu\nu} \qquad
\text{with} \\
T^{(\phi)}_{\mu\nu} & \equiv & L_\phi \, g_{\mu\nu} 
 +   \sum_{n=1}^p n  
 \left\{-2 \frac{\partial L_\phi}{\partial \chi_n} [\phi]^{n}_{\mu\nu} 
 +\nabla^\alpha \left[ \frac{\partial L_\phi}{\partial \chi_n}\left([\phi]^{n-1}_{\alpha\mu} \phi_\nu + 
 [\phi]^{n-1}_{\alpha\nu} \phi_\mu - [\phi]_{\mu\nu}^{n-1} \phi_\alpha\right)\right]\right\}\!\! .\label{stress}
\eea
To get rid of $\lambda$ in the Einstein equation, we proceed as in the previous case. First, we take the trace of the second equation in 
\eqref{general mimetic eom} to express $\lambda$ in terms of $\phi$ and $g_{\mu\nu}$
\bea
\lambda \; = \; \mp \frac{1}{2} ({\cal R}+T^{(\phi)}) \, , \quad
T^{(\phi)} \; = \; 4L_\phi - \sum_{n=1}^p n \, \left\{ 
2  \frac{\partial L_\phi}{\partial \chi_n} \chi_n + \nabla^\alpha \left[  \phi_\alpha \frac{\partial L_\phi}{\partial \chi_n} \chi_{n-1}  \right]
\right\}\, .
\eea
Then we substitute this expression in the two equations above \eqref{general mimetic eom}. Furthermore,
the equation for the scalar field can be obtained from the conservation of the stress-energy tensor, and thus is not
independent from Einstein equations. 

Hence, as the trace of Einstein equations is trivially satisfied (the trace has been used to determine $\lambda$), the equations of motion are
 equivalent to the mimetic condition and (the traceless part of) Einstein equations only:
\bea\label{mimetic equations}
X \pm 1 \; = \; 0 \, , \qquad G_{\mu\nu} \; = \;  \pm ({\cal R} + T^{(\phi)}) \, \phi_\mu \phi_\nu +  T^{(\phi)}_{\mu\nu}  \, .
\eea
Solutions to these equations have been studied in the context of cosmology \cite{Chamseddine:2016uef} and black holes
\cite{Chamseddine:2016ktu} with a particular choice for $L_\phi$ which makes the solutions nonsingular.  Here, we focus on 
black hole solutions and we are going to see how one can choose $L_\phi$ to resolve the black hole singularity.

\subsection{Black hole with a limiting curvature}
\label{Schar coord}
The non-singular  black hole introduced by Chamseddine and Mukhanov in \cite{Chamseddine:2016ktu} is a ``static" spherically symmetric solution of the general mimetic action \eqref{general mimetic action} where $L_\phi$ is a function of $\chi_1$ only defined by
\bea\label{CMmodel}
 L_\phi (\chi_1) \; = \; \frac{2}{3} \rho_m \, f(\zeta) \, , \quad \zeta = \frac{\chi_1}{\sqrt{\rho_m}} \, , \quad
f(\zeta) \equiv 1 + \frac{1}{2} \zeta^2- \sqrt{1-\zeta^2} - \zeta \arcsin \zeta \, ,
\eea
where $\rho_m$ defines a new energy scale in the theory. This expression of $L_\phi$ seems to be an ad hoc choice a priori, but it leads to very appealing nonsingular cosmological and
black hole solutions. Notice that, in the cosmological sector, the equation of motion of the scale factor  reproduces exactly the
 effective dynamics of LQC as it was pointed out in \cite{Liu:2017puc, Bodendorfer:2017bjt}.

\medskip

Let us now consider a spherically symmetric space-time only, and let us explain physically why 
\eqref{CMmodel}  produces non-singular black hole solutions. For that purpose, we start writing the metric
in Schwarzschild coordinates 
\bea\label{spherical}
ds^2 \; = \; -F(R)\, dT^2 + \frac{1}{F(R)} \, dR^2 + R^2\, (d\theta^2+ \sin^2\theta \, d\varphi^2) \, ,
\eea 
where $F$ is a function of $R$ only. In usual general relativity (with no modifications), Einstein equations lead to the  
Schwarzschild solution where $F(R)=1-2m/R$, $m$ being the mass of the black hole. Computing the expansion of outgoing null geodesics in the advanced Eddington-Finkelstein cordinates, non-singular when $F(R)=0$, one obtains $\theta_+=F(R)/R$. There exists thus a future trapping horizon at $R=2m$, which is an event horizon due to the staticity of the metric. One can therefore distinguish between the outside ($F>0$ or $R>2m$, thus $\theta_+>0$) and the inside of the black hole which is gravitationnally trapped ($F<0$ or $R<2m$, thus $\theta_+<0$). The singularity
occurs inside the black hole where the curvature becomes arbitrary large (in the limit $R \rightarrow 0$).

It was shown in the original paper  \cite{Chamseddine:2016ktu} that the action \eqref{general mimetic action} with the field Lagrangian \eqref{CMmodel} reproduces correctly the Schwarzschild metric far from the high curvature regions 
(compared with the scale $\rho_m$). In particular, spherically symmetric solution \eqref{spherical} possesses an event horizon, very similar
to the Schwarzschild horizon. Thus one can still define a region inside (or behind) the horizon ($F <0$) and a region outside the
horizon ($F>0$). 

Concerning the scalar field, let us start assuming that it depends on $R$ and $T$ for purposes of generality. We will shortly
reduce ourselves to the case of a static scalar field $\phi(R)$. The scalar field satisfies the mimetic condition which reads
\bea
-\frac{1}{F(R)} \left( \frac{\partial \phi}{\partial T}\right)^2 + F(R)  \left( \frac{\partial \phi}{\partial R}\right)^2 \pm  1 \; = \; 0 \, . 
\eea
This equation allows to resolve the scalar field $\phi$ in terms of the geometry $F(R)$.
A simple class of solutions of this partial differential equation can be obtained from the ansatz 
\bea
\phi(R,T)\, = \, qT+\psi(R) \, ,
\eea
 where $q$ is a constant and $\psi$
satisfies
\bea\label{eq1}
\left( \frac{d \psi}{d R}\right)^2 \; = \; \frac{q^2 \mp F}{F^2} \, .
\eea
Notice that similar ansatz were considered in \cite{Babichev:2016rlq,Babichev:2016fbg} to find black holes and stars solutions in the context of Horndeski (or beyond Horndeski) theories.

It is clear that the equation \eqref{eq1} admits a solution only if the condition $q^2 \mp F \geq 0$ is fulfilled. 
As a result, in the static case (where $q=0$), one cannot find any global spherically symmetric 
solution for the space-time. Indeed, the condition $\pm F \leq 0$ implies that only the action with a 
$+$ (resp. $-$) sign could lead to a description of the region inside (resp. outside) the black hole. 
Only a non-static solution for the scalar field ($q\neq 0$) could enable us to describe a fully static spherically
symmetric space-time. However, we will proceed as in  \cite{Chamseddine:2016ktu}: we will restrict ourselves to
the region inside the black hole (we expect the limiting curvature hypothesis to affect mainly the regions inside the black hole),
we choose a mimetic action with a $+$ sign, and we will argue how this is enough to 
resolve indeed the singularity. 
From a phenomenological point of view, we could interpret the action \eqref{general mimetic action}
with \eqref{CMmodel} as an effective description of general relativity in a region (inside the black hole) where the curvature becomes
high (with respect to the scale $\rho_m$). Such a modification could result from quantum gravity effects for instance \cite{Liu:2017puc}. 

When the scalar field is  static, the mimetic condition reduces to a simple differential  equation
\bea 
\left( \frac{d \phi}{d R}\right)^2 \; = \; -\frac{1}{F} \, ,
\eea 
in the region (behind the horizon) where $F\leq 0$ (with appropriate boundary conditions). 
The form of $L_\phi$ (the presence of $\arcsin(x)$ or $\sqrt{1-x^2}$ with $x=\chi_1/\sqrt{\rho_m}$ for instance) imposes
that the scalar field $\phi$ must satisfy the condition
\bea
\vert \chi_1 \vert \leq \sqrt{\rho_m} \; \Longrightarrow \; \vert \frac{d}{dR} \left( R^2 \sqrt{-F}\right) \vert \leq \sqrt{\rho_m} R^2  \, .
\eea
If one naively substitutes the Schwarzschild solution in this inequality, one gets the condition that
\bea
\label{BEnergyGen}
\rho \, \equiv \, \frac{m}{R^3} \; \leq \; \frac{2}{9} \, \rho_m \, ,
\eea  
which can be interpreted by the  fact that the density inside the black hole is bounded from above. Hence, one 
would expect the singularity to be resolved. This has been shown to be indeed the case in \cite{Chamseddine:2016ktu}
from a resolution of the equations of motion. We are going to reproduce this result in the next section from a Hamiltonian 
point of view.

\section{Hamiltonian description}
In this section, we perform the Hamiltonian analysis of the mimetic action 
with a limiting curvature. We first introduce the ADM parametrization for the metric and we solve the mimetic condition to integrate out
the scalar field $\phi$. Then, we start the Hamiltonian analysis and we find a nice parametrization of the phase space such that
the Hamiltonian and vectorial constraints take a rather simple  form. Finally, we resolve the Hamilton equations far behind the horizon,
and we recover that the solution is indeed non-singular. As we are going to see, the Hamiltonian point of view leads to a simpler analysis of 
the equations of motion than the Lagrangian
point of view, as it was done in \cite{Chamseddine:2016ktu}.

 \subsection{3+1 decomposition: metric and scalar field}
 In this subsection, we start introducing
 the tools  which are necessary to perform the Hamiltonian analysis of the theory restricted to spherically symmetric geometries.
 Notice that  the ``radial" and ``time" coordinates (in the Schwarzschild parametrization) exchange their roles when one crosses the horizon. 
In particular, the time coordinate behind the horizon would correspond to the radial coordinate in the Schwarzschild parametrization. 
 
\subsubsection{ADM decomposition for spherical space-time}
 We start with the usual ADM decomposition of the (non-static) spherically symmetric metric inside the black hole:
  \bea
 ds^2 \; = \; -N^2 dt^2 \, + \, \gamma_{rr}(dr+N^r dt)^2 \, + \, \gamma_{\theta\theta} d\Omega^2 \, , \quad
 d\Omega^2 \equiv  d\theta^2 + \sin^2\theta d\varphi^2\, ,
 \eea
 where $N(r,t)$ is the lapse function, $N^r(r,t)$ is the radial component of the shift vector and 
 \bea\label{spatialgamma}
\gamma \equiv \text{diag} \left[\gamma_{rr}(r,t),\gamma_{\theta\theta}(r,t), \gamma_{\theta\theta}(r,t)\sin^2\theta \right] 
 \eea
 are the non-vanishing components of the (spherically symmetric) induced metric on the three dimensional space-like hypersurface. 
In the following, we will use the standard notations $\gamma^{rr}\equiv \gamma_{rr}^{-1}$ and $\gamma^{\theta\theta}\equiv \gamma_{\theta\theta}^{-1}$ for the components of the inverse metric $\gamma^{-1}$.  
 
 \subsubsection{Resolution of the mimetic condition to integrate out the scalar field}
 Concerning the scalar field, we assume that it depends on time $t$ only (which corresponds to a ``static" solution from the point of view
 of an observer outside the horizon). The mimetic condition $X+1=0$ implies that the lapse function necessarily depends on $t$ according to
 \bea
 \dot{\phi}(t)^2 \; = \; N(t)^2 \, .
 \eea
 Without loss of generality, we take the solution $\dot \phi=+N$ that we substitute in the action in order to integrate out the
 scalar field  $\phi$. 
 To do so, we also need to compute $\chi_1=\Box \phi$ in terms of the metric variables. 
 
 Thus, we start by computing second 
 derivatives of the scalar field $\phi_{\mu\nu}$, and an easy calculation shows that the only non-vansihing  
 components of
 $\phi_{\mu\nu}$ are
 \bea\label{solmimetic}
 \phi_{tt} \; = \; - (N^r)^2 \, K_{rr} \, , \quad
  \phi_{rt} \; = \; -N^r \, K_{rr}  \, , \quad
 \phi_{ii} \; = \; - K_{ii} \, ,
 \eea
 where $i \in \{r,\theta,\varphi\}$ labels spatial coordinates, and $K_{ij}$ are the components of the extrinsic curvature
 \bea
 K_{ij} \; \equiv \; \frac{1}{2N}(\dot \gamma_{ij} - D_i N_j - D_j N_i) \, ,
 \eea
 with $D_i$ being the covariant derivative compatible with the spatial metric $\gamma_{ij}$ \eqref{spatialgamma}.
To go further, one has to compute explicitly the components of the extrinsic curvature. Only the diagonal components 
 are non-trivial with
\bea
K_{rr}=\frac{1}{2N} \left( \dot\gamma_{rr} - \gamma_{rr}' N^r - 2 \gamma_{rr} (N^r)'\right) \; , \quad
K_{\theta\theta}=\frac{K_{\varphi\varphi}}{\sin^2\theta} =\frac{1}{2N}\left( \dot{\gamma_{\theta\theta}} - \gamma_{\theta\theta}' N^r \right)\,.
\eea
As usual,  dot and prime denote respectively  time and radial derivatives. We have now all the ingredients to start the Hamiltonian analysis of the
theory.

\subsection{Hamiltonian analysis}
Using the Gauss-Codazzi relation, the action \eqref{mimetic action} with $f(\phi)=1$ and $L_\phi$ given by \eqref{CMmodel} reduces to
the form
\bea
S \; = \; \int dt \, d^3x \, N \sqrt{\gamma} \left[ \frac{1}{2} (K_{ij} K^{ij} - K^2 + R) + L_\phi(K) \right]\, ,
\eea
where we have substituted the solution of the mimetic constraint for the scalar field \eqref{solmimetic}, and $R$
is the three-curvature whose expression in terms of the metric components is
\bea
R \; = \; \frac{2}{\gamma_{\theta\theta}} \left[ 1- \left(\frac{\gamma_{\theta\theta}'}{\sqrt{\gamma_{\theta\theta}\gamma_{rr}}}\right)^2
+ \frac{2}{ \gamma_{rr}} \left( \frac{\gamma_{\theta\theta}'}{\sqrt{\gamma_{\theta\theta} \gamma_{rr}}}\right)' \right]\, .
\eea
To simplify the Hamiltonian analysis, it is more convenient to introduce a new parametrization of the metric in terms of the 
 variables $\alpha$ and $\gamma$ defined by
\bea
\label{Parmetrization}
\left\{\gamma \; \equiv \; \gamma_{rr} \gamma_{\theta\theta}^2 \sin^2{(\theta)} \; , \quad
\alpha \; \equiv \; \frac{\gamma_{rr}}{\gamma_{\theta\theta}} \right\} \, \Longleftrightarrow \;
\left\{ \gamma_{rr} = (\gamma \alpha^2)^{1/3} \; , \quad \gamma_{\theta\theta} = \left(\frac{\gamma}{\alpha}\right)^{1/3} \right\} \, .
 \eea
With these new variables,  the action simplifies and becomes
\bea\label{Lag alpha}
S & = & \int dt \, d^3x \,\sqrt{\gamma} \left[ \frac{1}{12N} \left( A^2 - B^2 \right) + N \left(  \frac{R}{2} + L_\phi \left( \frac{B}{2N}\right) \right)\right]
\eea
where $A$ and $B$ depend respectively on the variables $(\alpha,N^r)$ and $(\gamma,N^r)$ according to
 \bea
A \equiv \frac{\dot\alpha}{\alpha} - \frac{\alpha'}{\alpha} N^r - 2 (N^r)' \, , \qquad 
B \equiv \frac{\dot\gamma}{\gamma} - \frac{\gamma'}{\gamma} N^r - 2 (N^r)' \, .
\eea
Notice that $R$ can be expressed explicitly in terms of $\alpha$ and $\gamma$, what we will not do here because 
its explicit form is not needed for our purposes.  

It is clear from this expression of the Lagrangian \eqref{Lag alpha} that $N$ and $N^r$ are Lagrange multipliers, and there are only two pairs of
conjugate variables defined by the Poisson brackets
\bea\label{mimeticps}
\{ \alpha(u), \pi_\alpha(v)\} \; = \;  \delta(u-v) \; = \; \{ \gamma(u) , \pi_\gamma(v)\} \, .
\eea   
Momenta are easily computed and are given in terms of the velocities by
\bea
\pi_\alpha \;  = \; \frac{1}{3} \frac{\sqrt{\gamma}}{N\alpha} A \, , \qquad 
\pi_\gamma &  = & \, -\frac{1}{3} \sqrt{\frac{{\rho_m}}{{\gamma}}} \, \arcsin\left( \frac{B}{2N\sqrt{\rho_m}}\right) \, .
\eea
Inverting these relations to obtain velocities in terms of momenta is immediate
\bea
\frac{\dot \alpha}{\alpha} & = & 3 \frac{N\alpha}{\sqrt{\gamma}} \pi_\alpha + \frac{\alpha'}{\alpha} N^r + 2 (N^r)' \, , \\
\frac{\dot \gamma}{\gamma} & = & -2N\sqrt{\rho_m} \sin\left( 3\sqrt{\frac{{\rho_m}}{{\gamma}}} \pi_\gamma\right) + \frac{\gamma'}{\gamma} N^r + 2 (N^r)' \, .
\eea
From these expressions, one easily deduces the expression of the Hamiltonian 
\bea
H \; = \; \int dr \, \left( N {\cal H} + N^r D_r \right) 
\eea
with the Hamiltonian and vectorial constraints respectively given by
\bea\label{HHrmimetic}
{\cal H} &\equiv& \frac{3}{\sqrt{\gamma}} \alpha^2 \pi_\alpha^2 - 
\sqrt{\gamma} \left[ \frac{4}{3} \rho_m \sin^2 \left( \frac{3}{2} \sqrt{\frac{\gamma}{\rho_m}}\pi_\gamma\right) + \frac{1}{2}R\right]\, ,\\
 D_r &\equiv &-\left( \gamma' \pi_\gamma +2\gamma \pi_\gamma' + \alpha'\pi_\alpha + 2 \alpha \pi_\alpha'\right)\, .
\eea
 Hence, time derivative of any functionnal $\cal O$ of the phase space variables is computed from
the Poisson bracket
\bea
\dot{\cal O}(t,r) \; = \; \{ {\cal O} ; H \} \, ,
\eea
from which one easily deduces the equations of motion. To these equations, one adds the two constraints 
${\cal H} \approx 0$ and $D_r \approx 0$ to integrate completely the system. Notice that $\approx$ denotes the weak equality
in the phase space.

Notice that the Hamiltonian constraint can be written as 
\bea\label{rho}
\rho \equiv \frac{9\alpha^2}{4\gamma} \pi_\alpha^2 - \frac{3}{8} \, R \; \approx \; \rho_m \, 
\sin^2 \left( \frac{3}{2} \sqrt{\frac{\gamma}{\rho_m}}\pi_\gamma\right) \, ,
\eea
where  $\rho$ will be interpreted as the energy density of the mimetic scalar field, as we are going to argue later. 
The density $\rho$ is necessarily bounded by $\rho_m$. 

\subsection{Spherically ``static" solutions deeply inside the black hole}
To go further, we restrict ourselves to spherically static solutions which correspond, in the black hole interior, 
to having a time dependency only.

\subsubsection{Equations of motion}
 In that case, the vectorial constraint $D_r= 0$ is strongly
satisfied and the expression of the 3-dimensional Ricci scalar reduces to $R=2(\alpha/\gamma)^{1/3}$. 
Hence, after some simple calculations, one shows that the equations of motion for the phase space variables are 
\bea
&&\dot\alpha=\frac{6\alpha^2 \pi_\alpha}{\sqrt{\gamma}} \, , \quad
\dot\pi_\alpha = -\frac{6\alpha \pi_\alpha^2}{\sqrt{\gamma}}  + \frac{1}{3}\frac{\gamma^{1/6}}{ \alpha^{2/3}} \, , \quad
\dot\gamma = -2\gamma \sqrt{\rho_m} \, \sin \left( {3} \sqrt{\frac{\gamma}{\rho_m}}\pi_\gamma\right) \, , \label{eqgamma} \\
&&\dot\pi_\gamma = \frac{3}{2} \frac{\alpha^2\pi_\alpha^2}{\gamma^{2/3}} + \frac{\rho_m}{3\sqrt{\gamma}} 
\sin^2\left( \frac{3}{2} \sqrt{\frac{\gamma}{\rho_m}}\pi_\gamma\right) + \frac{\rho_m}{2} \pi_\gamma \sin\left( {3} \sqrt{\frac{\gamma}{\rho_m}}\pi_\gamma\right) - \frac{\alpha^{1/3}}{6 \gamma^{5/3}} \, ,
\eea
where we have fixed the lapse to the value $N=1$ for simplicity (which corresponds to a redefinition of the time variables). 
These equations are highly non-linear and very cumbersome to solve.
Even though we do not expect to find explicit solutions,  we are going to present  some interesting properties they satisfy. 

First of all, the equation for $\gamma$ \eqref{eqgamma} together with the Hamiltonian constraint \eqref{rho} leads to the so-called master 
equation in \cite{Chamseddine:2016ktu} given by
\bea\label{modified Fried}
\left(\frac{\dot\gamma}{4\gamma} \right)^2\; = \;  \,\rho\left( 1- \frac{\rho}{\rho_m}\right) \, , \quad \text{with} \quad
\rho \; = \; \frac{1}{4} \left[ \frac{1}{4} \left( \frac{\dot\alpha}{\alpha}\right)^2 - {3} \left(\frac{\alpha}{\gamma} \right)^{1/3} \right]\, .
\eea
In \cite{Chamseddine:2016ktu}, the same equation has been obtained from the Lagrangian point of view  in terms of the variable
$\epsilon \equiv (4/3) \rho$ and the constant $\epsilon_m \equiv (4/3) \rho_m$. 
We recover an equation very similar to the one satisfied by the scale factor in the framework of effective LQC \cite{Liu:2017puc}, 
which leads to a non-singular scenario for early universe cosmology. Here, we have a similar dynamics for $\gamma$ which is 
also bounded from below: this leads to a non-singular black hole solution \cite{Chamseddine:2016ktu} as it can be easily seen
from  the expression of the Kretschmann tensor  ${\cal K}=4(\alpha/\gamma)^{1/3}$: it is bounded when $\gamma$ does not vanish, provided that
$\alpha$ does not tend to infinity neither. 

\subsubsection{Deep inside the black hole: resolution of the singularity}
To understand better the dynamics of the variables $\alpha$ and $\gamma$, 
let us consider the regime where the 3-dimensional scalar curvature $R$ becomes 
negligible in the expression of the energy density \eqref{rho}. This hypothesis implies the condition
\bea\label{condition}
\frac{\alpha^2\pi_\alpha^2}{\gamma}  \gg R \quad \Longleftrightarrow \quad  \left(\frac{\dot\alpha}{\alpha} \right)^2 \gg 
\left(\frac{\alpha}{\gamma} \right)^{1/3} \, .
\eea
It has been shown in \cite{Chamseddine:2016ktu} that this condition is satisfied  deeply inside the black hole  (in the region where the quantum effects are supposed to be important). In this situation, the equations for $\alpha$ and $\pi_\alpha$ become
\bea
\dot\alpha=\frac{6\alpha^2 \pi_\alpha}{\sqrt{\gamma}} \, , \qquad
\dot\pi_\alpha = -\frac{6\alpha \pi_\alpha^2}{\sqrt{\gamma}} \, ,
\eea
which can be integrated exactly. First, $\alpha$ can be expressed in terms of $\gamma$ using the relations:
\bea\label{alpgamm}
\sqrt{\gamma} \, \frac{\dot \alpha}{\alpha} \; = \; C \quad  \Longleftrightarrow \quad
\alpha(t) \; = \; \alpha_0 \,  \exp \left( C \int_0^t \frac{du}{\sqrt{\gamma(u)}}\right) \, ,
\eea
where $C$ and $\alpha_0$ are integration constants. Then, the equation for $\gamma$ \eqref{modified Fried} simplifies and becomes
\bea
{\dot \gamma^2} \; = \; C^2 \left( \gamma - \gamma_m \right) \, , \qquad \gamma_m \equiv \frac{C^2}{16 \rho_m} \, ,
\eea
from which we deduce immediately that $\gamma$ is bounded from below  by $\gamma_m$. Furthermore, one easily integrates this
equation and obtains an exact expression for $\gamma$ (in the regime where $R$ is negligible in the expression of ${\cal H}_0$) given
by
\bea\label{gamma t}
\gamma(t) \; = \; \gamma_m (1+ 4 \rho_m t^2) \, ,
\eea
assuming that $\gamma(0)=\gamma_m$. This expression coincides completely with Eq.(64) of \cite{Chamseddine:2016ktu} 
when the constant $C$ has been fixed to $C=3r_g$ ($r_g$ being the Schwarzchild radius). Finally, substituting this expression
in \eqref{alpgamm}, one obtains 
\bea\label{alpha t}
\alpha(t) \; = \; a \, \exp \left[ 
2 \, \sinh^{-1} \left(2 \, \sqrt{\rho_m} t\right) 
\right] \, ,
\eea
where ${a}$ is a new constant. Hence, we recover exactly the solution found in \cite{Chamseddine:2016ktu} where
 $a$ has been fixed to
 \bea
 a \; = \; \frac{64}{9} \, \rho_m \, .
 \eea
 As a conclusion,
the singularity is clearly avoided and replaced by a bounce very similar to the LQC bounce. 

\medskip

Let us recall that the analysis of this subsection is valid only when the condition \eqref{condition} is fulfilled. Using
the relation \eqref{alpgamm}, the validity condition can be more explicitly given by
\bea
\label{condi}
C^6 \gg \alpha \, \gamma^2  \quad \Longleftrightarrow \quad
18^2 r_g^2 \rho_m \gg  (1+ 4 \rho_m t^2)^2 \, \exp \left[ 2 \, \sinh^{-1} \left(2 \, \sqrt{\rho_m} t\right) 
\right]  .
\eea
Thus,  $\rho_m t^2 \ll 1$ is  sufficient for the condition (\ref{condi}) to be satisfied provided that 
\bea
Q \; \equiv \; r_g^2 \rho_m \gg 1\, ,
\eea 
which is obviously the case deep inside the black hole. However, this is not necessary. Indeed, in the regime where $\rho_m t^2 \gg 1$, the previous condition gives
\bea\label{right cond}
\rho_m t^3 \, \ll \, {r_g} \, ,
\eea
which is less restrictive than $\rho_m t^2 \ll 1$. Hence, the solution (deep inside the black hole) given by \eqref{gamma t} and
\eqref{alpha t} is valid when $t$ is sufficiently small according to \eqref{right cond}. 

\subsubsection{Comparing with the Schwarzschild solution }
\begin{figure}
    \centering
    \includegraphics[width=7cm]{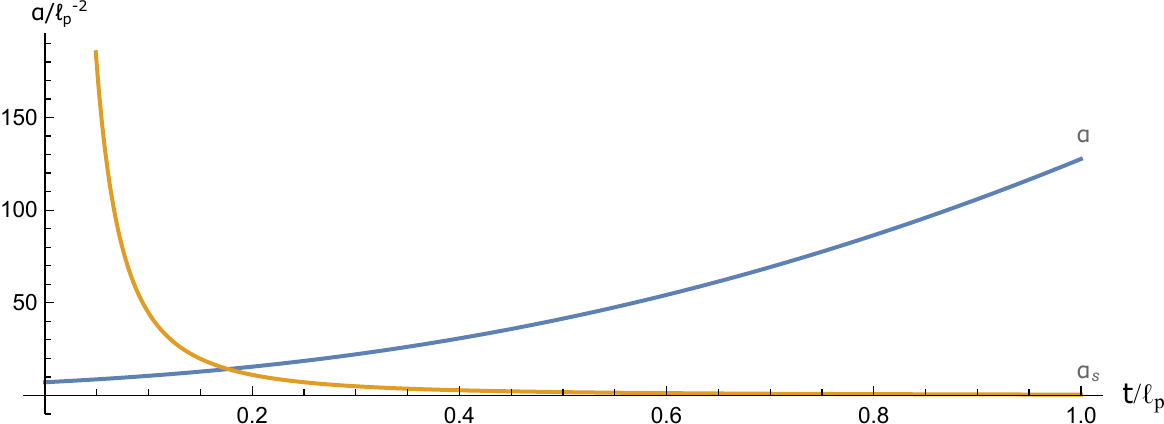}
    \includegraphics[width=7cm]{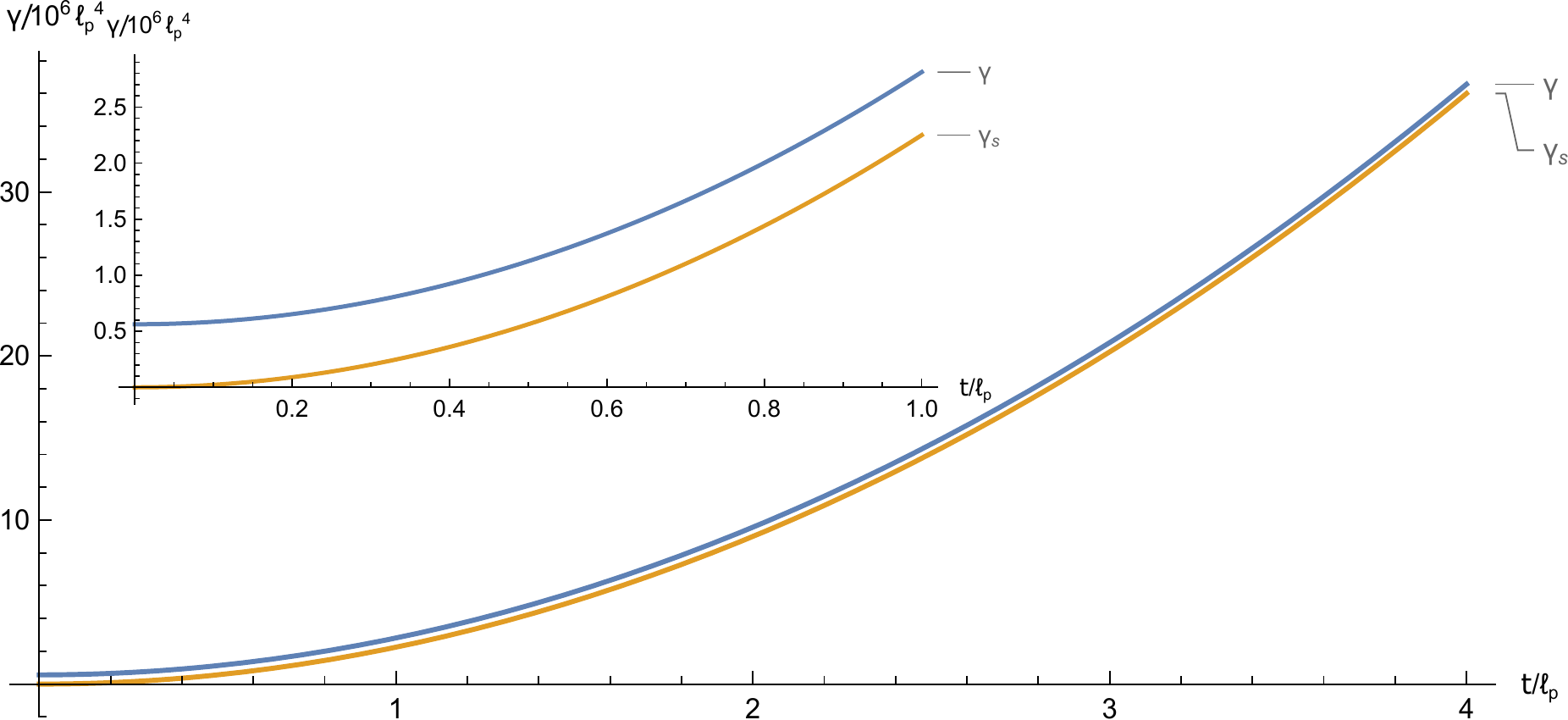}
    \caption{These two graphs show the solutions $\alpha(t)$ and $\gamma(t)$ compared to the Schwarzschild solution
    $\alpha_s(t)$ and $\gamma_s(t)$ in the deep quantum regime \eqref{right cond}. In particular, we see that the limiting curvature effect
    makes the functions $\alpha(t)$ and $1/\gamma(t)$ no more divergent when $t \rightarrow 0$, which regularizes the singularity. In these 
    plots, we work in the Planck unit with $r_g=1000 \ell_p$ and $\rho_m=\ell_p^{-2}$, where $\ell_p$ is the Planck length. In that case, the
     deep quantum regime condition  \eqref{right cond} becomes $t \ll 10 \ell_p$.}
    \label{fig:alphagamma}
\end{figure}

To conclude this analysis of the geometry deep inside the black hole, let us make a comparison with the usual Schwarzschild solution. 
When expressed in terms of the parameters $\alpha$ and $\gamma$, the Schwarzschild black hole is defined by 
(see \cite{Chamseddine:2016ktu} for instance)
\bea
\alpha_s \; = \; \frac{1-\tau^2}{r_g^2 \tau^6} \, , \quad
\gamma_s \; = \; r_g^4 (1-\tau^2)\tau^6  \quad \text{with} \quad
\frac{t}{r_g} \; \equiv \; \arcsin\tau - \tau\sqrt{1-\tau^2}  \, .
\eea
In the regime \eqref{right cond}, we necessarily have $t/r_g \ll Q^{-1/3} \ll 1$, hence $\tau \ll 1$ and then
\bea
\tau^3 \; \approx \; \frac{3t}{2r_g} \, . 
\eea
Thus, the Schwarzschild solution,
to compare the regularized solution with, reduces to
\bea
\alpha_s(t) \; \approx \; \frac{4}{9t^2} \, ,\quad
\gamma_s(t) \; \approx \; \frac{9}{4} r_g^2 t^2 \, ,
\eea
in the region where the curvature is high.
We see that both $\alpha$ and $\gamma^{-1}$ are regularized compared to the classical Schwarzschild solution as shown in 
Fig. \eqref{fig:alphagamma}: this is the effect
of the limiting curvature. However, close to the horizon (which corresponds to $\tau \approx 1$), one recovers the Schwarzschild
solution because the limiting curvature effect becomes negligible.
 
\section{On the relation with effective polymer black holes}

In \cite{Bodendorfer:2017bjt, Liu:2017puc}, it was shown that extended mimetic gravity with \eqref{CMmodel} reproduces the dynamics of effective LQC when restricted to homogenous and isotropic cosmology. A natural question is whether this relation extends itself to the spherically symmetric sector. We explore this question in this section where we compare the Hamiltonian of the non-singular black hole with limiting curvature  with the standard polymer model of vacuum spherically symmetric gravity in terms of Ashtekar-Barbero variables studied in \cite{Gambini:2013ooa}. 
While the relation between the Chamseddine-Mukhanov mimetic theory with limiting curvature and LQC is shown not to hold beyond homogenous geometries, one can propose new mimetic theories to describe a generalized regularization for polymer models which provides 
us with an undeformed notion of covariance as well as a natural inbuilt $\bar{\mu}$-scheme. 

\subsection{Spherically symmetric polymer models: a short review}

In this section, we first give a brief review on the formulation of the classical spherically symmetric Ashtekar-Barbero phase space and then we describe the so-called holonomy corrected phase space.
As explained in the introduction, this leads to an effective polymer model which has been the starting point for a quantization of spherically symmetric geometries using the LQG techniques \cite{Gambini:2013ooa}. 
This phase space, which describes vacuum spherically symmetric gravity, has only global degrees of freedom.  The quantum effective corrections are introduced via holonomy corrections which ensure the existence of an anomaly-free (albeit deformed) Dirac's hypersurface deformation algebra \cite{Tibrewala:2013kba, Bojowald:2015zha}. In that sense, the resulting polymer phase space describes a vacuum geometry with somehow a deformed notion of covariance.

\subsubsection{The  classical vacuum spherically symmetric Ashtekar-Barbero phase space}

Starting from the full Ashtekar-Barbero phase space, one can impose spherical symmetry to reduce it. This symmetry reduction has been presented in \cite{Bojowald:2004af} and we shall not reproduce the different steps in details but only sketch the main lines. 

The spherically symmetric Ashtekar-Barbero phase space is parametrized by two pairs of canonical conjugate variables
(after having gauge fixed the Gauss constraint) given by
\beq\label{eq:pb_lqg}
\{ K_x(x), E^x (y)\} = \frac{2 \kappa}{\beta} \,  \delta(x-y) \, ,\qquad \{ K_{\phi} (x) , E^{\phi} (y)\} = \frac{ \kappa}{\beta} \,  \delta(x-y) \, ,
\eeq
where $1/ \beta$ is the Barbero-Immirzi parameter and $\kappa=8\pi G$ with $G$ the newton constant. Note the $x$ and $y$ are radial 
coordinates and $\delta(x-y)$ denotes the usual one-dimensional delta distribution. 

The first pair corresponds to the inhomogenous component of the connection with its associated electric field, while the second one is the angular contribution built from the angular components of the connection.
As usual in the LQG litterature, $x$ denotes the radial direction as shown in the expression of the induced metric 
\beq\label{eq:met_lqg}
ds^2 \equiv {\gamma}_{rr} dr^2 + {\gamma}_{\theta\theta} d\Omega^2 \equiv \frac{(E^{\phi})^2}{|E^{x}|} dx^2 + |E^x| d\Omega^2 \, .
\eeq
The first class constraints generating the gauge symmetries (i.e. spatial diffeomorphisms and time reparametrization) are 
\begin{align}
& D[N^x] = \frac{1}{2\kappa} \int_{\Sigma} dx\;  N^x \left[ 2 E^{\phi} K'_{\phi} - K_x (E^x)' \right] \, ,\\
& {\cal H}[N] = - \frac{1}{2\kappa} \int_{\Sigma} dx \; N |E^x|^{-1/2} \left[E^{\phi} K^2_{\phi} + 2 K_{\phi} K_x E^x + (1 - \Gamma^2_{\phi}) E^{\phi} + 2 \Gamma'_{\phi} E^x \right] \, ,
\end{align}
where the spin-connection component which shows up in the expression of ${\cal H}$[N] is given by $\Gamma_{\phi} \equiv - (E^x)' / (2E^{\phi})$. The two constraints satisfy obviously the hypersurface deformation algebra of general relativity (after reduction to the spherically symmetric geometries)
\begin{align}
& \{ D[N^x], D[M^x] \} = D[ \cL_N M] \, ,\\
& \{ D[N^x], \cH[M]\} = - \cH [ \cL_N M] \, , \\
& \{ \cH[N], \cH[M]\} = D[\gamma^{xx} (N \partial_x M - M \partial_x N)] \, ,
\end{align}
where we used the same notations as in the equations \eqref{diffintro} and \eqref{Ham} in the introduction.

\subsubsection{The holonomy corrected polymer phase space}

Following the strategy used to study polymer models at the effective level,  one introduces holonomy corrections (called loop corrections) to work with holonomies instead of the connection itself. These effective corrections originate from a general regularization procedure borrowed from Loop Quantum Gravity\footnote{Note that the full regularization implies additional corrections: the triad corrections associated to the regularization of the inverse volume term. Yet, most of the investigations only implement one type of corrections while ignoring the second type for practical reasons. Even more, among the holonomy corrections, only the point wise quantum corrections are taken into account. Beyond the technical aspect, the motivation for focusing on the holonomy corrections only comes from the observation that in the cosmological sector, such corrections are enough to obtain a singularity resolution mechanism.}. However,  several new difficulties show up when considering an inhomogenous background such as vacuum spherically symmetric gravity compared to homogenous cosmological backgrounds. In particular, one has to ensure that the loop regularization does not generate anomalies in the first class constraints algebra. 

The usual regularization consists in making the replacement 
\beq
\label{StandardChoice}
K_{\phi}  \longrightarrow  f(K_{\phi}) \, ,
\eeq
directly in the expression of the Hamiltonian constraint, where the function $f$ encodes quantum gravity effects at the effective level. Such a procedure is obviously not unique and suffers from ambiguities. Furthermore, there could be as many different functions $f$ as 
many $K_\phi$ that appear in the Hamiltonian constraint. 
Yet, a standard choice (in the LQG litterature) is 
\bea
\label{sin}
f(K_{\phi}) \; = \; \frac{\sin{(\rho K_{\phi})}}{\rho} \, ,
\eea
where $\rho$ is a new fundamental (quantum gravity) scale, taken to be the minimal area gap for LQG,  such that  the limit $\rho \rightarrow 0$ reproduces the classical phase space. Notice that this choice is motivated by the polymerization obtained in LQC, from the computation of the non local curvature operator using $\SU(2)$ holonomies (one can refer to \cite{Ashtekar:2011ni} for more details). 

Moreover, the above effective corrections depend only on the extrinsic curvature while the parameter $\rho$, i.e. the so called polymer scale, is a constant. This type of correction is known in the terminology of LQC to correspond to the $\mu_0$-scheme. In the early stages of LQC, this scheme was used to develop the theory. Yet, it was soon realized that such corrections spoil the semi-classical consistency of the model. Indeed, within this $\mu_0$-scheme, the bounce of the universe as described by LQC could take place at any arbitrary energy density. The way out of this important difficulty was to introduce an improved dynamics based on a new type of effective holonomy corrections, the so called $\bar{\mu}$-scheme. In this new regularization, the polymer scale $\rho$ depends explicitly on the metric. Contrary to the $\mu_0$-scheme, where the semi-classical limit had to be taken by hand, i.e. $\rho \rightarrow 0$, the $\bar{\mu}$-scheme allows the polymer scale to run with the metric, and thus $\rho$ naturally vanishes in the large volume limit, as expected from a consistent regularization. 

A first attempt to introduce a $\bar{\mu}$-scheme in spherically symmetric loop models has been presented in \cite{Tibrewala:2012xb}. In that case, $f$ is now a function $f(K_{\phi}, E^x)$ of $K_\phi$ and $E^x$. It was shown that the resulting quantum corrections are not periodic anymore if one requires anomaly-freeness of the first class constraints algebra, challenging the possibility to interpret such  corrections as resulting fundamentally from $\SU(2)$ holonomy corrections.

With this standard regularization (\ref{StandardChoice}), the holonomy corrected first class constraints take the form
\begin{align}
\label{diff}
& D[N^x] = \frac{1}{2\kappa} \int_{\Sigma} dx\;  N^x \left[ 2 E^{\phi} K'_{\phi} - K_x (E^x)' \right] \, , \\
\label{scalLQG}
& \cH[N] = - \frac{1}{2\kappa} \int_{\Sigma} dx \; N |E^x|^{-1/2} \left[ E^{\phi} f_1( K_{\phi} ) + 2 f_2 ( K_{\phi}) K_x E^x + (1 - \Gamma^2_{\phi}) E^{\phi} + 2 \Gamma'_{\phi} E^x \right] \, ,
\end{align}
where the two functions $f_1$ and $f_2$ regularize the two $K_\phi$ arguments which appear in the expression of the classical Hamiltonian constraint. The corresponding Poisson algebra is
\begin{align}
& \{ D[N^x], D[M^x] \} = D[ \cL_N M] \, ,\\
& \{ D[N^x], \cH[M]\} = - \cH [ \cL_N M] \, , \\
& \{ \cH[N], \cH[M]\} = D[\beta{(K_{\phi})}\gamma^{xx} (N \partial_x M - M\partial_x N)] \, ,
\end{align}
 provided that the two corrections functions satisfy the differential equation
\beq
\label{A}
f_2 = \frac{1}{2}\frac{d f_1}{d K_{\phi}}   \, .
\eeq
This implies in turn that the deformation of the Dirac's algebra is given by
\beq
\label{deformation}
 \beta(K_{\phi}) = \frac{d f_2}{d K_{\phi}}  \, .
\eeq
 Interestingly, at the bounce, the sign of the function $\beta(K_{\phi})$ changes (when one takes the usual $\sin$ polymerization function (\ref{sin})) and the constraints algebra becomes effectively euclidean. This observation has suggested a possible signature change of the space-time in the very quantum region. This point is still debated and has received more attention in the context of the deformed algebra approach to the cosmological perturbations in LQC \cite{Grain:20100jlq, Grain:2016jlq}. See \cite{Bojowald:2015gra, Mielczarek:2014kea, JulienSignChange} for a general discussion on the conceptual and technical consequences of this deformation in early cosmology, and \cite{Bojowald:2014zla} in the context of black hole. Finally, note that the above deformation disappears when working with the self-dual Ashtekar variables. Indeed, thanks to the the self-dual formulation, one can introduce point wise holonomy corrections with a $\bar{\mu}$-scheme without affecting the Dirac's algebra which keeps its classical form. This was shown both for spherically symmetric gravity coupled to a scalar field and for the unpolarized Gowdy model, which both exhibit local degrees of freedom \cite{BenAchour:2016brs, BenAchour:2017jof}\footnote{See also \cite{BenAchour:2016leo} for similar conclusion in the inhomogeneous cosmological context.}.

In the case of a real Ashtekar-Barbero spherically symmetric polymer phase space, the usual choice for an effective description of spherically symmetric geometries in LQG corresponds to taking
\beq
 f_1(K_{\phi}) = \frac{\sin^2{(\rho K_{\phi})}}{\rho^2} \, , \qquad
 f_2(K_{\phi}) = \frac{\sin{(2\rho K_{\phi})}}{2\rho} \, ,
\eeq
which obviously satisfies  \eqref{A}.

\subsubsection{Comparing mimetic and polymer black holes}

Now, we compare the Hamiltonian structures of LQG and mimetic gravity with a limiting curvature. Following the previous section, we
can write the effective constraints of spherically symmetric LQG as follows:
\begin{align}
& \cH =  |E^x|^{-1/2} \left[ E^{\phi} \frac{\sin^2{(\rho K_{\phi})}}{\rho^2} + 2 \frac{\sin{(2\rho K_{\phi})}}{2\rho} K_x E^x + (1 - \Gamma^2_{\phi}) E^{\phi} + 2 \Gamma'_{\phi} E^x \right] \, ,\\
& D_x =   2 E^{\phi} K'_{\phi} - K_x (E^x)'  \, .
\end{align}

To make a comparison with mimetic gravity in the spherical sector, we need to reformulate the mimetic phase space \eqref{mimeticps} 
and the mimetic constraints \eqref{HHrmimetic} in terms of the LQG variables. 
Let us start with the phase space variables. First of all, the coordinates $\alpha$ and $\gamma$ are easily related to the components
$E^\phi$ and $E^x$ of the electric field as follows: 
\begin{equation}
\label{coord}
\alpha={\gamma}_{rr} {\gamma}_{\theta \theta}^{-1}=(E^{\phi})^2 (E^x)^{-2} \, , \quad \gamma={\gamma}_{rr} {\gamma}_{\theta \theta}^{2}=(E^{\phi})^2 E^x \, .
\end{equation}
To obtain the relation between the momenta ($\pi_\alpha,\pi_\gamma$) and the components ($K_\phi,K_x$) of the extrinsic curvature, we
transform the symplectic potential $\Theta$ of the LQG phase space
\bea
\Theta \; \equiv \; \beta \left[K_{\phi} \delta E^{\phi} +\frac{1}{2}{K_{x}} \delta {E^x}\right] \, ,
\eea
as follows
\bea
\Theta & = &\beta \left[K_{\phi} \delta ( \alpha^{\frac{1}{6}} \gamma^{\frac{1}{3}}) +\frac{1}{2}K_x \delta ( \alpha^{-\frac{1}{3}} \gamma^{\frac{1}{3}}) \right] \, , \\
&= &\beta \left[ \frac{K_{\phi}}{3 E^{\phi} E^x } + \frac{K_{x}}{6 (E^{\phi})^2 } \right] \delta {\gamma} + \beta \left[ \frac{K_{\phi} (E^x)^2 }{6 E^{\phi} } - \frac{K_{x} (E^x)^3}{6 (E^{\phi})^2 } \right] \delta {\alpha} \, .
\eea
Due to the equality $\Theta = \pi_{\gamma} \delta \gamma + \pi_{\alpha} \delta \alpha$,
we deduce  immediately the useful relations
\bea
\label{mom}
\pi_{\alpha}  \, = \,  \beta \left[ \frac{K_{\phi} (E^x)^2 }{6 E^{\phi} } - \frac{K_{x} (E^x)^3}{6 (E^{\phi})^2 } \right] \, , \quad
\pi_{\gamma} \, =   \, \beta \left[ \frac{K_{\phi}}{3 E^{\phi} E^x } + \frac{K_{x}}{6 (E^{\phi})^2 } \right] \, .
\eea
Substituting \eqref{coord} and \eqref{mom} in \eqref{HHrmimetic}, one immediately obtains the expressions of the mimetic Hamiltonian constraint in terms of the LQG variables
\begin{align}
\cH = \frac{\beta^2}{12} \frac{(K_{\phi}E^{\phi}-K_x E^x)^2}{E^{\phi}\sqrt{E^x}}-\frac{4\rho_m}{3} E^{\phi}\sqrt{E^x} 
 \sin^2 \bigg{(} \frac{\beta}{6} \frac{2 K_{\phi}E^{\phi}+K_x E^x}{ \sqrt{\rho_m} E^{\phi}\sqrt{E^x} } \bigg{)} -\frac{1}{2} E^{\phi}\sqrt{E^x} R \, .
\end{align}
Hence, the Hamiltonian constraints of mimetic gravity and LQG do not  not coincide. The identification we have noticed for homogeneous and isotropic backgrounds does not extend to spherically symmetric geometries. 
 While only the variable $ K_{\phi}$ is polymerized in the polymer Hamiltonian (using a constant scale $\rho$), it is a rather complicated combination of the different fields that enters in the sine function in mimetic gravity, namely
\beq
\label{QPOLY}
\cA = \tilde{\rho} \big{(} K_{\phi}  +  \frac{E^x}{E^{\phi}}K_{x} \big{)} \qquad \text{with} \qquad \tilde{\rho} =  \frac{\beta}{3 \sqrt{\rho_m E^x}} \, .
\eeq
Notice that in this last quantity, $\tilde{\rho}$ depends on the phase space field $E^x$, and provides a natural $\bar{\mu}$-scheme. 

The result obtained in this section is not surprising. Indeed, since the effective polymer model has a deformed notion of covariance compared with General Relativity, one does not expect to find a covariant effective action which would reproduce its Hamiltonian formulation. However, one can use mimetic gravity with a limiting curvature as a guide to build undeformed covariant notion of point-wise holonomy like corrections. Therefore, while we do not expect to reproduce any existing inhomogeneous polymer models from these scalar-tensor theories (albeit in the cosmological sector where the issue of covariance disappears), we want to build new extended mimetic Lagrangians which reproduce ``as closely as possible" the Hamiltonian of the current spherically symmetric polymer models. The difference between the mimetic and polymer Hamiltonian formulations will then provide an interesting guide to understand the absence of covariance in inhomogeneous polymer black holes. The next section is devoted to this task.

\subsection{Polymer Hamiltonians from extended mimetic gravity}
In this section, we present a method such that, given effective corrections introduced at the hamiltonian level in the spherically symmetric phase space of vacuum gravity, one can reconstruct the associated extended mimetic covariant Lagrangian. The theory remains fully covariant despite the effective hamiltonian corrections and thus extended mimetic gravity provides us with a guide to build a covariant regularization of the classical phase space of spherically symmetric gravity.
\subsubsection{General construction}
Our starting point is again the most general form for the extended mimetic action \eqref{general mimetic action}
\bea
\label{GA}
S[g_{\mu\nu},\phi,\lambda] \; \equiv \; \int d^4x \, \sqrt{-g} \, \left[ \frac{f(\phi)}{2} \, {\cal R} + 
\, L_\phi(\phi,\chi_1,\cdots,\chi_p) \, + \, \lambda( X \pm 1)\right]\, ,
\eea
where $f$ is an arbitrary function of $\phi$, and $L_\phi$ depends on $\phi$ and $\chi_n$ 
defined in  \eqref{chin}. For simplicity, we choose $f=1$ and we assume that $L_\phi$ does not depend on $\phi$. 

Thanks to the mimetic constraint, in the ``static" case where the scalar field depends on time $t$ only, one can relate directly the second order derivative of the scalar field $\phi_{\mu\nu}$ to the extrinsic curvature of the 3-hypersurface $\Sigma$ of the ADM decomposition. Using (\ref{solmimetic}), one can rewrite the variable $\chi_n$ as
\begin{equation}\label{eq:chi_K}
      \chi_n =(-1)^{n}\sum_{i_1,...,i_n} K_{i_1}^{i_2}K_{i_2}^{i_3} ... K_{i_{n-1}}^{i_n}K_{i_n}^{i_1} \,.
 \end{equation}
In the spherically symmetric case, $K_{r}^{r}$, $K_{\theta}^{\theta}$ and $K_\varphi^\varphi$ ($\propto K_{\theta}^{\theta}$) are the only non-trivial components of $K_\mu^\nu$. In this simplified context, 
it is more convenient to introduce the following combinations of these components 
\begin{align}{\label{eq:xy_met}}
& X=K_{r}^{r}+K_{\theta}^{\theta} =\frac{\dot{E}^{\phi}-(N^r {{E}^{\phi}})'}{NE^{\phi}}, &Y=K_{\theta}^{\theta}=\frac{\dot{E}^{x}-N^r {{E}^{x}}'}{2 N E^x} \, ,
\end{align} 
which involve separately the velocities of $E^\phi$ and $E^x$. This will be very useful for the Hamiltonian analysis.  Their covariant form can be given in terms of $\chi_1=\Box \phi$ and $\chi_2= \phi_{\mu\nu} \phi^{\mu\nu}$ only as follows
\begin{align}\label{eq:x_y}
&X=\frac{2}{3} \Box \phi + \frac{1}{6} \sqrt{6{\phi_{\mu \nu} \phi^{\mu \nu}}-2{(\Box \phi)}^2} \, ,
&Y=\frac{1}{3} \Box \phi - \frac{1}{6} \sqrt{6{\phi_{\mu \nu} \phi^{\mu \nu}}-2{(\Box \phi)}^2} \, .
\end{align} 

Hence, we can view the $\chi_i$ variables as functions of only the two variables $X$ and $Y$, and then we can  reformulate 
the general action (\ref{GA}) as
\begin{equation}
\label{GAA}
S=\int dt d^3 x \; N \; E^{\phi}\sqrt{E^x} \left[ -(2 XY-Y^2) + \tilde{L}_{\phi}(X,Y)+\frac{1}{2} R \right] \, ,
\end{equation}
where $\tilde{L}_{\phi}(X,Y)=L_\phi(\chi_1,\cdots,\chi_p)$. This  expression of the action is much more suitable for a Hamiltonian analysis. First, we compute the momenta conjugated to the variables $E^{x}$ and $E^{\phi}$ 
\bea
\pi_x & = & \frac{\delta Y}{\delta \dot{E}^{x}}\frac{\delta L}{\delta Y} =\frac{E^{\phi}}{2 \sqrt{E^x}} (-2 (X+Y) + \frac{\partial \tilde L_{\phi}}{\partial Y} ) \, ,\label{Momenta1}\\ 
\pi_{\phi} & = & \frac{\delta X}{\delta \dot{E}^{\phi}}\frac{\delta L}{\delta X} = \sqrt{E^x} (-2 Y + \frac{\partial \tilde L_{\phi}}{\partial X} ) \, ,\label{Momenta2}
\eea
where $L$ denotes the full Lagrangian of the theory. Notice that in the classical limit, where ${\tilde{L}}_{\phi}(X,Y) \rightarrow 0$, the momenta become ``classical" and reduce to
the expected form
\beq
\pi_x \rightarrow \pi^c_x  = - \frac{E^{\phi}}{ \sqrt{E^x}}  (X+Y) \quad \text{and} \quad 
\pi_\phi \rightarrow \pi^c_{\phi} = - 2 \sqrt{E^x}  Y \, .
\eeq
To compute the Hamiltonian, one has to invert the relations \eqref{Momenta1} and \eqref{Momenta2} in order to express the velocities 
in terms of the momenta. However, it is not always possible to have an explicit inversion and then to have an explicit form of the
Hamiltonian. 

Our goal is to find a function ${\tilde{L}}_{\phi}(X,Y)$ which has the right semi-classical limit and which reproduces the general shape of the quantum correction appearing in the polymer model as shown in (\ref{scalLQG}). Since such corrections only affect the angular component of the extrinsic curvature in the polymer model, we focus on $\pi_\phi$ in \eqref{Momenta2}, and we require that there
exists a function $f$ such that
\bea
Y=- {f \left(\frac{\pi_{\phi}}{\sqrt{E^x}}\right) } \, .
\eea
Remark that such a quantum correction has a natural inbuilt $\bar{\mu}$-scheme through its dependency in $E^x$ to keep the covariance. 
Since the function $f$ does not depend on $X$, we can conclude that the general function ${\tilde{L}}_{\phi}(X,Y)$ which reproduces  the polymer Hamiltonian takes the form:
\begin{equation}\label{eq:lp_g}
\tilde{L}_{\phi} (X, Y)=2 X Y + X f^{-1} (-Y) + g(Y) \, ,
\end{equation}
where $g(Y)$ is a function of $Y$  only which satisfies $g(Y) \rightarrow 0$ in the classical limit. 
Indeed, this choice of $\tilde{L}_{\phi}(X,Y)$ leads to the action (\ref{GAA}) 
\begin{align}
\label{GAAA}
S & =\int dt d^3 x \; N \; E^{\phi}\sqrt{E^x} \left[ \; Y^2 + X f^{-1} (Y)+ g(Y) +\frac{1}{2} R \;  \right] \, ,
 \end{align}
 whose Hamiltonian (after a few calculations) reads
 \begin{align}
 H & = N \left[2  \ E^x \; \pi_{x} \;  Y +  E^{\phi}\;  \pi_{\phi}\; X -   \; E^{\phi}\sqrt{E^x} \left( \; Y^2 + X f^{-1} (Y) + g(Y) +\frac{1}{2} R \; \right)\right] \notag \\
& = - \frac{N}{\sqrt{E^x}} \; \left[2 E^x \pi_x f \left(\frac{\pi_{\phi}}{\sqrt{E^x}} \right) + E^{\phi} \left( f^2 \left(\frac{\pi_{\phi}}{\sqrt{E^x}} \right) - E^x g(Y) \right) + \frac{1}{2} E^{\phi} E^x R \right]\\
 &=- \beta^2 \frac{N}{\sqrt{E^x}} \; \left[\frac{2}{\beta} K_x E^x f \left(\frac{2 \beta K_{\phi}}{\sqrt{E^x}} \right) + E^{\phi} \left( \frac{1}{\beta^2}f^2 \left(\frac{2\beta K_{\phi}}{\sqrt{E^x}} \right) - E^x g(Y) \right) + \frac{1}{2\beta^2} E^{\phi} E^x R \right] \, ,\notag
 \end{align}
 where we have used the relations \eqref{mom}.
 Notice that one can express the three-curvature $R$ as in (\ref{scalLQG}) but this is not needed here. 
 As a conclusion, one can immediately identify this Hamiltonian with the  polymer model Hamiltonian (\ref{scalLQG}) with the conditions that
 \beq\label{eq:f1_f2_f}
 f_1 (E^x, K_{\phi}) =   \frac{1}{\beta^2} f^2 \left(\frac{2\beta K_{\phi}}{\sqrt{E^x}} \right) - E^x g(Y) \qquad \text{and} \qquad f_2 (E^x, K_{\phi}) = 
 \frac{1}{\beta} f \left(\frac{2 \beta K_{\phi}}{\sqrt{E^x}} \right) \, .
 \eeq
 In that way, one obtains an extended theory of mimetic gravity which reproduces the general shape of the holonomy corrections considered in the polymer model, with an additional natural $\bar{\mu}$-scheme.

\medskip

Let us now quickly look at the cosmological limit of the previous action. In that case, all the $\chi_i$ variables  can be expressed in terms of $\Box \phi$ only, and we have
\begin{align}\label{eq:lqc_lim_xy}
X= 2 Y = \frac{2}{3} \Box \phi \, .
\end{align}
As a consequence,  the mimetic potential takes the very simple form 
\begin{equation}
L_{\phi}= \frac{4}{9}\left(\Box \phi \right)^2 + \frac{2}{3} \Box \phi \, {f}^{-1} \left(\frac{2}{3} \Box \phi\right) +g \left(\frac{1}{3} \Box \phi\right) \, .
\end{equation}

Having presented our general procedure to derive an effective action reproducing the hamiltonian effective corrections of polymer models, let us briefly summarize our strategy. The general mimetic action we shall investigate in the following, through special examples, is given by
\bea
\label{GA}
S[g_{\mu\nu},\phi,\lambda] \; \equiv \; \int d^4x \, \sqrt{-g} \, \left[ \frac{1}{2} \, {\cal R} + 2 X Y + X f^{-1} (-Y) + g(Y) 
  \, + \, \lambda( \phi_{\mu}\phi^{\mu} \pm 1)\right] \;\;\;\;\;
\eea
where the variabes $X$ and $Y$ are given by
\bea
X=\frac{2}{3} \Box \phi + \frac{1}{6} \sqrt{6{\phi_{\mu \nu} \phi^{\mu \nu}}-2{(\Box \phi)}^2} \, \qquad Y=\frac{1}{3} \Box \phi - \frac{1}{6} \sqrt{6{\phi_{\mu \nu} \phi^{\mu \nu}}-2{(\Box \phi)}^2} \, .
\eea
and the functions $f^{-1} (-Y)$ and $g(Y)$ are the unknown.

We can now apply our strategy to concrete examples treated in the polymer litterature. An interesting question that we intend to clarify with the next two examples is the following. Considering the two loop regularization proposed so far in spherically symmetric inhomogeneous backgrounds, does their associated effective mimetic actions derived from the procedure outlined above coincide with the Chamseddine-Muhkanov's lagrangian as the cosmological lagrangian ? We show below that in order to have the right cosmological limit, the anomaly free condition within the $\bar{\mu}$-scheme are a crucial ingredient. 

\subsubsection{A first example: the standard anomaly free sine function corrections}

As a first example, let us consider the standard holonomy corrections given by sine functions which satisfy the anomaly free constraint (\ref{A}):
\begin{align}
\label{first}
f_1 (K_{\phi})= \frac{\sin( \rho K_{\phi})}{\rho} \,, \quad  f_2(K_{\phi})=\frac{1}{2} \frac{d f_1(K_{\phi})}{d K_{\phi}} \, .
\end{align}
Here the scale $\rho$ is a constant and thus, does not correspond to the $\bar{\mu}$-scheme. As we have seen, when deriving the quantum corrections from the covariant theory, we always obtain a dynamical scale $\rho$ which depend explicitly on the metric. Hence, the corrections (\ref{first}) cannot be reproduced as they stand from our procedure. 

A first naive attempt is to generalize the above correction in the $\mu_0$ scheme to the $\bar{\mu}$ one, by promoting the scale $\rho$ to a dynamical scale $\tilde{\rho} = 2 \rho \beta /\sqrt{E^x}$. This generalization is not harmless from the point of view of the anomaly freedom problem. Indeed, when generalizing the corrections function in this way, we should first derive the new associated anomaly free conditions which will be different from (\ref{first}) and then compute the solutions to this conditions (See example 2 below). Here, we proceed to the generalization by hand and show, as expected, the covariant mimetic lagrangian which reproduces these naive $\bar{\mu}$ corrections does not admit the cosmological Chamseddine-Mukhanov's lagrangian at its cosmological limit. This first example underlines how the anomaly-free conditions combined with a $\bar{\mu}$-scheme are crucial for the consistency of the model. 

After our naive generalization to include by hand a $\bar{\mu}$-scheme, our corrections functions become
\begin{equation}
\label{HC}
f_1 (E^x, K_{\phi})= \frac{\sqrt{E^x}}{2 \beta \rho} \sin\left( \frac{2 \beta \rho}{\sqrt{E^x}} K_{\phi}\right), \;\;\;\;\;
 f_2 (E^x, K_{\phi})=\frac{E^x }{ \beta^2 \rho^2}  \sin^2\left( \frac{\beta \rho}{\sqrt{E^x}} K_{\phi}\right)\, .
\end{equation} 

From the previous construction, we can explicitly derive the mimetic Lagrangian which reproduces such holonomy-like corrections (\ref{HC}).
Using (\ref{eq:f1_f2_f}), one can first write the angular component of the extrinsic curvature $Y = - K^{\theta}_{\theta}$ as
follows
\begin{equation}
\begin{split}
Y=-f \Big(\frac{2 \beta K_{\phi}}{\sqrt{E^x}} \Big)=-\beta f_2(E^x, K_{\phi}) \, .
\end{split}
\end{equation}
Then, one easily obtains the expressions of the functions $f$ and $g$ entering in the definition of $\tilde L_\phi$
\bea
	f^{-1}(-Y) = \frac{1}{\rho} \arcsin (-2 \rho Y )\, , \qquad
	g(Y)=-Y^2 + \frac{1}{2 \rho^2} (1 - \sqrt{1- 4 \rho^2 Y^2}) \, ,
\eea
which finally  leads to the  Lagrangian
\begin{equation}
\tilde L_{\phi} =2XY-Y^2 +\frac{X}{\rho} \arcsin \left(-2 \rho Y \right)+ \frac{1}{2 \rho^2} \left(1 - \sqrt{1- 4 \rho^2 Y^2}\right)\, .
\end{equation}
The symmetry reduction  to the cosmological sector leads
\begin{equation}
\label{LA}
L_{\phi}=\frac{1}{3} \Box \phi^2 - \frac{2}{3\rho} \Box \phi \arcsin \left(\frac{2}{3} \rho \Box \phi \right)+ \frac{1}{2 \rho^2} \left(1 - \sqrt{1- \frac{4}{9} \rho^2 \Box \phi^2}\right) \, .
\end{equation} 
As expected,  (\ref{LA}) does not coincide with the extended mimetic Lagrangian initially proposed by Chamseddine and Mukhanov (\ref{CMmodel}) which has been shown to  reproduce the LQC dynamics \cite{Liu:2017puc,Bodendorfer:2017bjt}. 
Therefore, the naive generalization of the effective corrections (\ref{first}) to (\ref{HC}) does admit a covariant mimetic lagrangian albeit with the wrong cosmological limit.  The reason is that the corrections (\ref{HC}) does not satisfy the rigth anomaly free conditions. We can now turn to the second example which exhibits the right anomaly free conditions. 

\subsubsection{A second example: the Tibrewala's effective corrections}

As emphasized above, the corrections functions (\ref{HC}) satisfy the anomaly free conditions (\ref{first}) which were derived for corrections functions $f_{1,2}(K_{\phi})$. Yet such conditions are not consistent with a $\bar{\mu}$-scheme. Such conditions for corrections functions of the type $f_{1,2}(E^x, K_{\phi})$ were derived in \cite{Tibrewala:2012xb} and read 
\begin{align}
\label{condition}
f_2 - f_1 \frac{\partial f_1}{\partial K_{\phi}} + 2 E^x \frac{\partial f_2}{\partial E^{x}} = 0
\end{align}
Choosing the function $f_2$ to be 
\begin{align}
\label{T1}
	f_2 (K_{\phi})= \frac{\sqrt{E^x} \sin( 2 \beta \rho K_{\phi}/\sqrt{E^x})}{2 \beta \rho}
\end{align}
the second holonomy corrections $f_1$ is constrained to be
\begin{align}
	\label{T2}
	f_1 (K_{\phi})= \frac{3 E^x \sin^2(\beta \rho K_{\phi}/\sqrt{E^x})}{\beta^2 \rho^2}-2 K_{\phi} \frac{\sqrt{E^x} \sin( 2 \beta \rho K_{\phi}/\sqrt{E^x})}{2 \beta \rho}
\end{align}
Following exactly the same strategy as in the previous example, we first find the functions $f$ and $g$ which are given here by
\begin{align}
	&f^{-1}(-Y) = \frac{1}{\rho} \arcsin (-2 \rho Y ) \, ,\\
	&g(Y)= \frac{3}{2 \rho^2} (1 - \sqrt{1- 4 \rho^2 Y^2})-Y^2-\frac{\arcsin(2 \rho Y)}{\rho} Y \, .
\end{align}
Then, we deduce immediately the mimetic Lagrangian 
\begin{equation}
\label{TibrewalaLag}
	\tilde L_{\phi}=2XY-Y^2 +\frac{X+Y}{\rho} \arcsin (-2 \rho Y )+ \frac{3}{2 \rho^2} (1 - \sqrt{1- 4 \rho^2 Y^2}) \, .
\end{equation}
Interestingly, the symmetry reduction to the cosmological sector leads to the Lagrangian
\begin{equation}
 	L_{\phi}=\frac{1}{3} \Box \phi^2 - \frac{1}{\rho} \Box \phi \arcsin (\frac{2}{3} \rho \Box \phi )+ \frac{3}{2 \rho^2} (1 - \sqrt{1- \frac{4}{9} \rho^2 \Box \phi^2})
\end{equation} 
which coincides exactly with the Chamseddine-Mukhanov Lagrangian, the one that reproduces the LQC dynamics (\ref{CMmodel}). Hence, one can see that the conditions (\ref{condition}), which are consistent with a $\bar{\mu}$-scheme, allow to derive a covariant mimetic lagrangian (in the spherically symetric sector) which admits the right cosmological limit, ie the lagrangian reproducing the LQC effective dynamics. The reason behind this result lies in the fact that the corrections (\ref{T2}) (derived in \cite{Tibrewala:2012xb}) do satisfy consistent anomaly free conditions which take into account, from the start, the existence of a $\bar{\mu}$-scheme. 

Yet, let us emphasize a subtelty of our work. The regularized phase space obtained from either implementing the effective corrections at the hamiltonian level as done in \cite{Tibrewala:2012xb} and the one obtained from a canonical analysis of the action (\ref{TibrewalaLag}) derived in this section, \textit{are not the same}. The first one has a deformed notion of covariance, the second exhibits an undeformed covariance since it is derived from a covariant action. This suggests that additional ingredients for the loop regularization of  \cite{Tibrewala:2012xb} to provide an undeformed covariance are encoded in the mimetic covariant action. From this observation, extended mimetic gravity provides indeed an interesting guide to investigate further the covariance issue of the loop regularization performed in polymer black hole and the relation between the two phases spaces should be studied further. We leave this observation for future work. 

Finally, we point out that from this investigation, the $\bar{\mu}$-scheme seems to be a necessary ingredient, albeit not sufficient, to have a fully covariant regularization of the phase space of spherically symmetric vacuum gravity and thus, hope to derive effective covariant actions for such polymer models\footnote{Note also that, using our procedure, it is possible to derive an effective action for the polymer black hole models which describe the Schwarzschild interior \cite{Ashtekar:2005qt, Corichi:2015xia}, since they are dealing with an anisotropic, albeit homogeneous background. In these models (at least the most recent version), a $\bar{\mu}$-scheme is included, just as standard LQC, and it is straitforward to derive the associated effective mimetic action reproducing their hamiltonian corrections. Yet the true challenge is dealing with inhomogeneous backgrounds, and thus we do not present the effective mimetic action for these polymer black hole interior models.}.

\section{Conclusion}
In this paper, we start by revisiting the regular black hole interior solution obtained in the context of mimetic gravity \cite{Chamseddine:2016ktu} from a Hamiltonian perspective. We introduced a suitable parametrization of the static spherically symmetric metric
which  allowed us to perform a complete Hamiltonian analysis of the theory. We wrote the Hamiltonian equations and we 
showed that the determinant $\gamma$ of the spatial metric admits an evolution equation very similar to the modified Friedmann equation obtained in effective LQC (\ref{modified Fried}). Thus, the black hole has a bounded energy density which prevents the existence of a 
singularity. Finally, we solved the Hamilton equations in the regime deep inside the black hole and we made a comparison with the standard
Schwarzschild solution, showing explicitly that the limiting curvature mechanism indeed resolves the singularity.

Then, motivated by the recent finding that the initial extended mimetic theory reproduces exactly the effective dynamics of LQC, we have investigated whether this result survives in the spherically symmetric sector. We have argued that this cannot be 
``naively" the case for a very general reason:  the spherically symmetric polymer model fails to be covariant, and there is a priori no hope to reproduce this phase space and its effective quantum corrections from a covariant theory such as mimetic gravity.  
From a more general perspective, the polymer treatment of inhomogeneous backgrounds (cosmological perturbations, spherical symmetry, Gowdy system) is well known to suffer from a lack of covariance. The holonomy corrections usually introduced during the regularization break the Dirac's hypersurface deformation algebra, or at best, lead to its deformation \cite{Bojowald:2015zha, Bojowald:2015sta, Bojowald:2016hgh, Bojowald:2016vlj, Bojowald:2016itl}. It becomes then crucial to understand either the conceptual and technical consequences of such deformation, either to understand how to cure it. In this work we have presented a new strategy towards the second problem based on mimetic gravity. Since this theory provides a covariant Lagrangian which reproduces exactly the effective LQC dynamics in the cosmological sector, it provides an interesting tool to derive covariant polymer-like Hamiltonian models beyond the cosmological sector to be compared with the existing ones. As such, we view mimetic theory as a guide to derive covariant notion of point-wise holonomy-like corrections in polymer models. 

With this issue of covariance in mind, we have presented a general procedure to construct mimetic Lagrangians which admit a Hamiltonian formulation very similar in spirit to existing polymer models of black holes, but which is fully covariant. Then we have applied our procedure to two examples of polymers black hole models: the standard anomaly-free sine corrections model and the Tribrewala's $\bar{\mu}$-scheme corrections model. From the differences between the mimetic and polymer Hamiltonian formulations, one can extract several insights which could be useful when trying to build undeformed covariant polymer black hole models based on the real Ashtekar-Barbero variables. 
First, the covariant mimetic formulation always contains a $\bar{\mu}$-scheme, i.e. holonomy-like corrections which depend both on $K_{\phi}$ and $E^x$ and not only on $K_{\phi}$. As such, the dynamical nature of the polymer scale $\rho$ is a necessary ingredient to maintain the covariance in presence of effective quantum corrections of the polymer type. However, the $\bar{\mu}$-scheme, while necessary, is not sufficient since previous attempts to include it in spherically symmetric models still lead to a deformed notion of covariance \cite{Tibrewala:2012xb} (see \cite{Han:2017wmt} for a similar conclusion in the context of cosmological perturbations). Therefore additional ingredients are required to maintain the covariance in such effective polymer models.
The second difference between the two approaches concerns the object which is polymerized. In the standard polymer approach to the spherically symmetric background, one polymerizes the connection $K_{\phi}$ while the limiting curvature mechanism of mimetic theory suggests that the covariance requires to polymerize a more complicated combination of the canonical variables, given by (\ref{QPOLY}). The choice of connection is crucial in this procedure, and the Ashetkar-Barbero connection $K_{\phi}$ might not be the suitable one to consider\footnote{It is already known that this connection does not transform as a true space-time connection under the action of the scalar constraint, and as such, does not provide a good candidate to ensure a fully covariant quantum description when implementing the dynamics. It might be that the same problem  emerges already at the effective level in these polymer models.}.
%The last difference between the two Hamiltonian formulations is the presence of the function $g(Y)$ within the mimetic Hamiltonian which is not there in the polymer one. When trying to mimic the effective corrections (\ref{HC}), or (\ref{T1} - \ref{T2}), this additional contribution is never vanishing. In the second example, which is the only polymer model including a $\bar{\mu}$-scheme, the mimetic and the polymer Hamiltonians do not coincide precisely because of the non vanishing contribution $g(Y)$ on the mimetic side.  Understanding how this contribution could be generated within the polymer framework seems to be a first step in order to build a covariant notion of such polymer models.

From a more general point of view, our study suggests that the current procedure developed in spherically symmetric polymer models still lacks some crucial ingredients to provide a consistent covariant effective framework.
However, let us emphasize again that the polymer models we have considered only involve the minimal loop corrections, i.e. the so called point-wise holonomy corrections. Therefore, it is still possible to improve  construction of polymer effective actions by adding the loop corrections. It could well be that the so called triad corrections, associated to the loop regularization of the inverse volume term, conspire with the holonomy corrections to provide an undeformed algebra of first class constraints. This remains to be checked. Moreover, the spherically symmetric sector raises the question of how to concretely deal with the holonomy of the inhomogeneous component of the connection $K_x$. Despite some preliminary work on this question, a consistent and tractable implementation of this extended holonomy is still to be understood. Therefore, our results might suggest that polymer models built on solely point-wise holonomy corrections (and with the real Ashtekar-Barbero variables) could not admit covariant effective action beyond the cosmological sector, and motivate us to look for a generalization of the current regularization used in such polymer models to implement consistently an undeformed notion of covariance (see \cite{Campiglia:2016fzp, BenAchour:2017qpb} for some recent proposals in black hole and cosmology in this direction). The recent exact effective black hole metric solutions for polymer black holes obtained in \cite{BenAchour:2018khr, Bojowald:2018xxu} show indeed that a generalization of the loop regularization to include a consistent $\bar{\mu}$ scheme is crucial to obtain black hole backgrounds with a consistent semi-classical limit. The present work suggests that extended mimetic gravity could provide an interesting guide towards this goal.

\begin{acknowledgments}
  HL thanks the hospitality of the Department of Physics, Florida Atlantic University, where some of the research related to this work was carried out. J.BA would like to thank the Theory Group of APC for its hospitality during his stay in Paris where this project was initiated. 
\end{acknowledgments}


\end{document}